\documentclass[twocolumn]{aa}  % for a paper on 2 column  

\usepackage{txfonts}
\usepackage{graphicx}
\usepackage{epsfig}
\usepackage{t1enc}                            
\usepackage{longtable}                     

\begin{document}

% TITLE:
   \title{The temporal changes of the pulsational periods of the pre-white dwarf PG~1159-035}

% AUTHORS:
   \author{J. E. S. Costa\inst{1}
           \and S. O. Kepler\inst{1}
          }

% OFFPRINTS:
   \offprints{00112740@ufrgs.br}   % <--- EMAIL...

% INSTITUTIONS:
   \institute{Instituto de F\'{\i}sica, Universidade Federal do Rio Grande do Sul, 
	91501-970 Porto Alegre, RS, Brazil% 1
%       \email{edu@if.ufrgs.br}
   }

% DATE:
   \date{Received ---; accepted 16/07/2008}

% ABSTRACT:
   \abstract {}{}{}{}{}  

   \abstract
  % Context (optional).
     {\object{PG~1159-035}, a pre-white dwarf with $T_{\mathrm{eff}}\simeq 140\,000$~K,
       	is the prototype of the PG~1159 spectroscopic class and  the DOV pulsating class. 
	Pulsating pre-white dwarf stars evolve rapidly: 
        the effective surface temperature decreases rapidly, the envelope contracts and 
        the inner structure experiences stratification due to gravitational settling.
	These changes in the star generate  variations in its oscillation periods. 
	The measurement of temporal change in the oscillation periods, $\dot P$, 
        allows us to estimate directly
	rates of stellar evolutionary changes, such as the cooling rate 
        and the  envelope contraction rate,
	providing a way to test and refine evolutionary models for 
	pre-white dwarf pulsating stars.
     }
  % Aims (mandatory).
     {  Previously, only two pulsation modes of the highest amplitudes for PG~1159-035 
	have had their $\dot P$ measured: the 516.0 s and the 539.3 s modes. 
       	  We measured the $\dot P$ of a larger number of pulsation modes, increasing the number
          of constraints for evolutionary studies of PG~1159-035.
          We attempted to use the secular variations in
          the periods of multiplets to calculate the variation in the rotational period, 
          the envelope contraction rate, and the cooling rate of the star.
     }
  % Methods (mandatory)
     { The period variations were measured directly from the PG~1159-035 observational data
          and refined by the (O-C) method.
     }
  % Results (mandatory).
     { We measured 27 pulsation mode period changes.
	The periods varied at rates of between 1 and 100 ms/yr, and several
	can be directly measured with a relative standard uncertainty  
        below 10\%. For the 516.0 s mode
	(the highest in amplitude) in particular,
        not only the value of $\dot P$ can be measured directly with 
        a relative standard uncertainty of 2\%, 
        but the second order period change, $\ddot P$, can also be calculated reliably.
	By using the (O-C) method, we refined the $\dot P$s and estimated the $\ddot P$s
        for six other pulsation periods.
	As a first application, we calculated the change in the PG~1559-035 rotation period,
	${\dot P}_{\rm rot}=(-2.13\pm 0.05)\times 10^{-6}\,{\rm ss}^{-1}$,
	the envelope contraction rate $\dot R=(-2.2\pm 0.5)\times 10^{-13}\,{\rm R}_\odot\rm{s}^{-1}$,
        and the cooling rate $\dot T=-1.42\times 10^{-3} {\rm Ks}^{-1}$.
     }
  % Conclusions (optional).
     {}

% KEYWORDS:
   \authorrunning{Costa and Kepler}
   \titlerunning{The changes of the pulsational periods in PG~1159-035}
   \maketitle

%------------------------------------------------------------------------
\section{Introduction}   % Sect. 1
%------------------------------------------------------------------------

%%%
\object{PG~1159-035} is the prototype of two star classes: the PG1159 spectroscopic
class and the DOV pulsating class, discovered by McGraw et al. (1979). 
The estimated temperature for PG~1159-035 is $140\,000\pm 5\,000$ K 
(Werner et al. 1991; Dreizler et al. 1998; Jahn et al. 2007), which  
places this star in the pre-white dwarf sequence of the Hertzsprung-Russell diagram. 

%%%
Analyzing the light curve obtained by the {\it Whole Earth Telescope} (WET) (Nather et al. 1990) 
in the 1989 campaign, Winget et al. (1991) (henceforth W91) identified 122 pulsation modes  in PG~1159-035 with 
periods between 300 and 1000 seconds, and  spherical harmonic index $\ell=1$ or $\ell=2$.
The excellent results allowed W91 to calculate the asteroseismological mass of the star
($M/M_\odot \simeq 0.59$)  and several other stellar parameters, such as the rotation period 
($P_{\rm rot} = 1.38\pm 0.01$ days), the inclination of the rotation axis ($i \simeq 60^o$), 
and a limit to the magnetic field strength ($B< 6000$ G). 
Costa et al. (2008), analyzing the PG~1159-035  combined data sets from  1983, 1985, 1989, 1993, and 2002,
increased the number of identified pulsation modes to 198, the largest after the Sun, 
refining the determination of the rotation period ($P_{\rm rot} = 1.3920\pm 0.0008$ days) and other stellar 
parameters previously estimated by W91. 
The comparison of the Fourier transforms of light curves from different years indicates that the amplitudes 
of the pulsation modes change with time, 
and their intensities in the light curves can decrease to lower than the lower limit for reliable detection.
Of the 198 detected modes, about 75\% appear in only one of the FTs, about 25\%   in two or more FTs,
and only 14 of the 198 pulsation modes  (all of them with index $\ell=1$) appear in the five FTs.

%%%
About $97\%$ of all stars end their evolution as white dwarfs. 
The pre-white dwarf stage is one of the ``front doors'' to the white dwarf cooling sequence. 
There are two evolutionary channels that are known to occur prior to the white dwarf stage.
The first channel involves the evolution of the star from the Horizontal Branch to white dwarf 
by means of  the Asymptotic Giant Branch (AGB), the Planetary Nebula phase, and PG1159 type stars.
In the second channel, the star evolves directly from the Extended Horizontal Branch to the white dwarf stage,
but without passing through  the Planetary Nebula phase. 

%%%
In its evolution, a pre-white dwarf star can traverse the DOV (PG1159 or {\it GW Vir}) instability strip, 
becoming a multiperiodic non-radial pulsating star, of periods between 100 and 1000 seconds. 
The exact values of the pulsation periods are defined by the mass, temperature, 
rotation, magnetic field, and structural characteristics of the star. 
While it passes through the DOV/PNNV instability strip, the star cools, its envelope
contracts, and its interior experiences stratification due to the gravitational settling.
As a consequence, the pulsation periods change with time.
The hotter the star is, the quicker are the changes in its pulsation periods.
For example, in \object{G117-B15A}, a DAV white dwarf star with $T_{\rm eff}\simeq 12,000$ K, 
the highest amplitude period, 215 s, changes 1 second in 8 million  years,
while in the hotter DOV star, PG~11559-035, with $T_{\rm eff}\simeq 140,000$ K, 
the 516 s period, the one of the highest amplitude, changes 1 second in only 350 years.

%%%
This work is a complementary study of the results obtained in Costa et al. (2008).
We used the measured periods in the PG~1159-035 light curves of the 1983-2002
data sets to calculate the secular variation in  several pulsation modes of PG~1159-035,
enlarging the number of constraints for future evolutionary studies of this star. 
As an immediate application, we show in Sect.~\ref{sect:6} 
how the measured $\dot P$s in a multiplet can be used to calculate 
the cooling rate of the star, the envelope contraction rate and the variation in the rotational period.
Part of the procedure used here was proposed by 
Kawaler et al. (1985a), Kawaler (1986) and other authors, 
but was never applied to (pre-)white dwarf stars 
because this requires precise period and $\dot P$ determinations. 
As highlighted in Sect.~\ref{sect:6}, the calculations are based on many
hypothesis about the interior of the star,  
but it is a first approach taking into account the  available observational data.
PG 1159-035 is the first and only white dwarf in which we can apply these theories
and test their limitations and reliability.

%%%
% TABLE 1  ------------------------------------
{\scriptsize
\begin{table*}
\caption{Observed periods (in seconds) in each yearly data set.} 
\label{table:1}
{\scriptsize
 \begin{center}
	% Tabela com os perídos para cada ano.
\begin{tabular}{ccrcccccc} \hline\hline
Mode  & $\ell$ & $m$ & $k\pm 1$ & 1983 & 1985 & 1989 & 1993 & 2002 \\
 (s)  &        &     &     &      &      &        &    &      \\ \hline
      &        &     &     &      &      &        &    &      \\ 

 376.0   &  2 &  0 & 22 &   $376.6472 \pm 0.0019$ & $376.0411 \pm 0.0057$ &   ---  &   ---  &   ---  \\
 387.4   &  2 &  0 & 27 &      ---  &   ---  & $387.4972 \pm 0.0164$ &   ---  & $387.4878 \pm 0.0160$ \\
 390.3   &  1 & -1 & 14 &   $390.2763 \pm 0.0009$ &   ---  & $390.2959 \pm 0.0050$ &   ---  & $390.3486 \pm 0.0316$ \\
 390.8   &  1 &  0 & 14 &     ---  &   ---  &   ---  & $390.7740 \pm 0.0152$ & $390.9327 \pm 0.0476$ \\
 397.2   &  2 & +2 & 28 &     ---  & $397.2251 \pm 0.0042$ &   ---  &   ---  & $397.2591 \pm 0.0226$ \\
 399.0   &  2 & +1 & 28 &     ---  &   ---  & $398.9141 \pm 0.0140$ &   ---  & $399.1425 \pm 0.0112$ \\
 400.0   &  2 &  0 & 28 &     ---  &   ---  & $400.0585 \pm 0.0039$ & $400.0396 \pm 0.0057$ & $400.0383 \pm 0.0121$ \\
 401.6   &  2 & -2 & 28 &     ---  &   ---  &   ---  &   ---  & $401.1103 \pm 0.0201$ \\
 412.0   &  1 &  0 & 15 &     ---  &   ---  & $412.0097 \pm 0.0093$ &   ---  & $412.0451 \pm 0.0122$ \\
 414.3   &  2 & -1 & 28 &     ---  &   ---  & $414.3894 \pm 0.0093$ & $414.4157 \pm 0.0086$ & $414.3356 \pm 0.0114$ \\
 415.5   &  2 & -2 & 28 &     ---  & $415.4274 \pm 0.0030$ & $415.6016 \pm 0.0050$ &   ---  & $415.7585 \pm 0.0155$ \\
 422.5   &  2 & +2 & 30 &   $422.5397 \pm 0.0008$ &   ---  &   ---  & $422.5630 \pm 0.0027$ & $422.5768 \pm 0.0056$ \\
 426.2   &  2 & -1 & 30 &     ---  &   ---  &   ---  & $426.2398 \pm 0.0064$ & $426.3075 \pm 0.0089$ \\
 427.5   &  2 & -2 & 30 &     ---  &   ---  & $427.5585 \pm 0.0033$ & $427.5125 \pm 0.0060$ & $427.4328 \pm 0.0514$ \\
 430.3   &  1 & +1 & 16 &   $430.2347 \pm 0.0032$ &   ---  &   ---  &   ---  & $430.3800 \pm 0.0407$ \\
 434.2   &  1 & -1 & 16 &     ---  & $434.3305 \pm 0.0030$ &   ---  &   ---  & $434.1443 \pm 0.0179$ \\
 436.5   &  2 & +1 & 31 &     ---  &   ---  & $436.5258 \pm 0.0160$ & $436.5766 \pm 0.0074$ & $436.5366 \pm 0.0226$ \\
 439.2   &  2 & -1 & 31 &     ---  &   ---  & $439.2814 \pm 0.0134$ & $439.2268 \pm 0.0089$ & $439.2376 \pm 0.0149$ \\
 440.6   &  2 & -2 & 31 &     ---  &   ---  & $440.6920 \pm 0.0061$ & $440.6430 \pm 0.0043$ & $440.4827 \pm 0.0124$ \\
 451.5   &  1 & +1 & 17 &   $451.5928 \pm 0.0005$ & $451.5776 \pm 0.0012$ & $451.5906 \pm 0.0019$ & $451.6007 \pm 0.0015$ & $451.5806 \pm 0.0030$ \\
 452.4   &  1 &  0 & 17 &   $452.4275 \pm 0.0008$ & $452.3713 \pm 0.0022$ & $452.4310 \pm 0.0021$ & $452.4324 \pm 0.0022$ & $452.4478 \pm 0.0035$ \\
 453.2   &  1 & -1 & 17 &   $453.2534 \pm 0.0016$ &   ---  &   ---  & $453.2800 \pm 0.0039$ & $453.2743 \pm 0.0066$ \\
 493.7   &  1 & +1 & 19 &   $493.7640 \pm 0.0005$ & $493.7490 \pm 0.0013$ & $493.7984 \pm 0.0024$ & $493.7941 \pm 0.0025$ &   ---  \\
 494.8   &  1 &  0 & 19 &   $494.9043 \pm 0.0012$ &   ---  & $494.8633 \pm 0.0178$ & $494.8046 \pm 0.0094$ &   ---  \\
 511.9   &  2 &  0 & 37 &     ---  &   ---  & $512.0018 \pm 0.0293$ & $511.9741 \pm 0.0123$ &   ---  \\
 515.0   &  2 & -1 & 37 &  $515.0718 \pm 0.0031$ & $515.0311 \pm 0.0033$ &   ---  & $515.0300 \pm 0.0140$ & $514.9390 \pm 0.0194$ \\
 516.0   &  1 & +1 & 20 &  $516.0260 \pm 0.0004$ & $516.0374 \pm 0.0006$ & $516.0548 \pm 0.0016$ & $516.0663 \pm 0.0014$ & $516.1028 \pm 0.0019$ \\
 517.1   &  1 &  0 & 20 &   $517.1225 \pm 0.0023$ & $517.1402 \pm 0.0014$ & $517.1663 \pm 0.0024$ & $517.1827 \pm 0.0020$ &   ---  \\
 518.2   &  1 & -1 & 20 &     ---  &   ---  & $518.2884 \pm 0.0028$ & $518.2960 \pm 0.0027$ & $518.2980 \pm 0.0077$ \\
 526.4   &  2 & -1 & 38 &    ---  &   ---  & $526.3592 \pm 0.0566$ &   ---  & $526.4515 \pm 0.0165$ \\
 536.9   &  1 & +1 & 21 &   $536.8412 \pm 0.0031$ &   ---  & $536.9530 \pm 0.0075$ & $536.8223 \pm 0.0155$ & $537.0081 \pm 0.0127$ \\
 538.1   &  1 &  0 & 21 &   $538.1585 \pm 0.0008$ &   ---  & $538.1547 \pm 0.0026$ &   ---  & $538.1696 \pm 0.0047$ \\
 539.3   &  1 & -1 & 21 &   $539.3593 \pm 0.0006$ & $539.4030 \pm 0.0010$ & $539.3552 \pm 0.0020$ & $539.3546 \pm 0.0012$ &   ---  \\
 540.9   &  2 & -2 & 39 &     ---  &   ---  &   ---  & $540.9620 \pm 0.0150$ & $540.8784 \pm 0.0600$ \\
 544.3   &  2 & +2 & 40 &     ---  &   ---  &   ---  & $544.3258 \pm 0.0204$ & $544.3305 \pm 0.0235$ \\
 557.1   &  1 & +1 & 22 &     ---  &   ---  & $557.1386 \pm 0.0034$ & $557.1137 \pm 0.0138$ & $557.1383 \pm 0.0139$ \\
 558.4   &  1 &  0 & 22 &     ---  &   ---  & $558.4476 \pm 0.0043$ & $558.4449 \pm 0.0073$ & $558.4297 \pm 0.0066$ \\
 559.7   &  1 & -1 & 22 &     ---  &   ---  & $559.7135 \pm 0.0124$ & $559.7608 \pm 0.0078$ & $559.7461 \pm 0.0258$ \\
 561.9   &  2 &  0 & 41 &   $561.9714 \pm 0.0028$ &   ---  &   ---  & $562.0109 \pm 0.0226$ & $561.7732 \pm 0.0261$ \\
 641.4   &  1 &  0 & 26 &     ---  &   ---  & $641.5343 \pm 0.0138$ & $641.4278 \pm 0.0329$ &   ---  \\
 644.9   &  1 & -1 & 26 &     ---  & $644.8953 \pm 0.0043$ & $644.9868 \pm 0.0165$ &   ---  &   ---  \\
 668.5   &  1 &  0 & 27 &     ---  &   ---  & $668.5441 \pm 0.0446$ &   ---  & $668.4789 \pm 0.0590$ \\
 685.8   &  1 & +1 & 28 &     ---  &   ---  & $685.7817 \pm 0.0899$ & $685.8593 \pm 0.0337$ &   ---  \\
 689.7   &  1 & -1 & 28 &    ---  &   ---  & $689.7606 \pm 0.0321$ & $689.8163 \pm 0.0234$ &   ---  \\
 705.8   &  1 & +1 & 29 &     ---  & $705.8474 \pm 0.0094$ & $705.9310 \pm 0.0220$ &   ---  &   ---  \\
 727.0   &  1 & +1 & 30 &     ---  &   ---  & $727.1008 \pm 0.0250$ &   ---  & $727.0057 \pm 0.0425$ \\
 729.6   &  1 &  0 & 30 &     ---  &   ---  & $729.5007 \pm 0.0614$ & $729.7205 \pm 0.0409$ &   ---  \\
 731.6   &  1 & -1 & 30 &     ---  &   ---  & $731.4527 \pm 0.0187$ & $731.6075 \pm 0.0280$ & $731.7440 \pm 0.0706$ \\
 755.3   &  1 & -1 & 31 &     ---  &   ---  & $755.3732 \pm 0.0635$ & $755.2315 \pm 0.0489$ &   ---  \\
 812.5   &  1 & +1 & 34 &     ---  & $812.4437 \pm 0.0124$ & $812.5725 \pm 0.0537$ &   ---  &   ---  \\
 819.7   &  2 & -2 & 60 &   $819.7720 \pm 0.0049$ &   ---  & $819.9484 \pm 0.0277$ & $819.5586 \pm 0.0289$ &   ---  \\
 842.8   &  1 & -1 & 35 &   $842.8731 \pm 0.0032$ &   ---  & $842.8873 \pm 0.0258$ &   ---  &   ---  \\
 861.7   &  1 &  0 & 36 &     ---  & $861.6775 \pm 0.0092$ &   ---  & $861.8524 \pm 0.0293$ &   ---  \\
 877.6   &  1 & +1 & 37 &     ---  &   ---  &   ---  & $877.6142 \pm 0.0703$ & $877.7300 \pm 0.0660$ \\

         &    &    &    &      &      &      &      &      \\ \hline
 \end{tabular}

 \end{center}
}
\end{table*}
}
% -------------------------------------------

%%%
% ------------------------------------------------------------------------
\section{Measurement of secular changes in period}  % Sect.2
% ------------------------------------------------------------------------

%%%
% --------------------------------------------------------------
\subsection{Period as a function of time}     % Sect. 2.1 
%%%
If we assume that a pulsation period $P$ is a continuous
and smooth function of time, $P=P(t)$, then it can be expanded  in terms of a
Taylor series,

\begin{eqnarray}
  P(t-t_o)  & = &     P(t_o) + \left. {dP\over{dt}}\right|_{t=t_o} (t-t_o) 
                    + \left. {1\over 2} {d^2P\over{dt^2}}\right|_{t=t_o} (t-t_o)^2 \\ \nonumber
            & + &   \left. {1\over 6} {d^3P\over{dt^3}}\right|_{t=t_o} (t-t_o)^3 
                    + \cdots
\label{eq:1} 
\end{eqnarray}

\noindent
where $t_o$ is an instant of time used as reference and $d^nP/dt^n$ are the period derivatives 
of $n^{\rm th}$ order. 
In the above equation, 
the number of terms that {\it must} be considered depends on the 
values of the derivatives and the time interval $\Delta t=t-t_o$ covered by the observational data. 
For DAV pulsating stars, which evolve more slowly, the terms  higher than  first 
order can be neglected, but for DOV stars (and even some DBV stars), 
it is necessary to consider the second order term,

\begin{equation}
  P(t-t_o) \simeq P_o + {\dot P_o}\,(t-t_o) + \frac{1}{2}\, {\ddot P_o}\,(t-t_o)^2\quad ,
 \label{eq:2}  % 
\end{equation}
where $P_o=P(t_o)\,$ and $\dot P_o$ and $\ddot P_o$ are the first and second order derivatives,
respectively, calculated at $t=t_o$,

\begin{equation}
 \dot P_o \equiv \left. {dP\over{dt}}\right|_{t=t_o}\quad \quad \mbox{ and}
 \label{eq:3}
\end{equation}

\begin{equation}
 \ddot P_o \equiv \left. {d^2P\over{dt^2}}\right|_{t=t_o}\quad .
 \label{eq:4} 
\end{equation}

% ------------------------------------------------------------------------------
\subsection{Direct measurement of $\dot P$}  % Sect. 2.2

%%%
If a period $P$ changes linearly  during the time interval $\Delta t$ ($\ddot P\simeq 0$), 
the $\dot P$ accuracy, $\sigma_{\dot P}$, considering two measurements only, can be estimated by

\begin{equation}
    \sigma_{\dot P} \simeq \sqrt{2}\, \frac{\sigma_P}{\Delta t} \quad ,
    \label{eq:5}  %%\label{eq_sigPdot} 
\end{equation}
where $\sigma_P$ is the uncertainty in the period determination. 
For PG~1159-035, $\sigma_P$ has values of
between $\sim 0.002$ and $0.2$ seconds. For $\Delta t = 19$ yr, we are able to
measure directly $\dot P$  over $\sim 3\times 10^{-12}\, {\rm ss}^{-1}$  for
the most accurately determined periods and over $\sim 1.6\times 10^{-10}\,{\rm ss}^{-1}$ for the others.
In other words, with the 19-year dataset it is not possible to measure directly $\dot P$ 
below $\sim 3\times 10^{-12}\,{\rm ss}^{-1}$,
even for the most accurately measured periods.
However, we can use the (O-C) method discussed in Sect.~\ref{sect_omc} 
to estimate $\dot P$ below this limit and 
improve all results obtained by direct measurement.

% --------------------------------------------------------------------------
\subsection{The (O-C) method\label{sect_omc} }   % Sect. 2.3

%%%
The (O-C) method (see e.g. Kepler 1993) 
is the most well-established method for the measurement of          
secular changes in pulsating star periods. 
Practically, its efficiency has been demonstrated by studies of several pulsating  
stars, such as the DAV star G117-B15A (Kepler et al. 2005a; Kepler et al. 2005b) 
and the DBV star R548 (Mukadam et al. 2003).  

%%%
By definition, the period of a periodic signal is the derivative of  the time, $t$, 
relative to the cycle fraction (epoch), $E$:
\begin{equation}
  P\equiv \frac{dt}{dE} \quad .
  \label{eq:6}
\end{equation}
From this definition,  we can derive an equation to determine the instant of time at which a maximum
in the pulsation cycle occurs, $T_{\rm max}$, in terms of a power series of the cycle number,
\begin{equation}
    T_{\rm max} = T_o + P_o\cdot E + \frac{1}{2}\cdot P_o \cdot {\dot P}\cdot E^2
                     + \frac{1}{6} \left({\ddot P}\cdot {P_o}^2 + {\dot P}^2\cdot P_o\right)\cdot E^3
                     + \cdots
    \label{eq:7}                 %%%\label{eq_c3} 
\end{equation}
where $T_o$ is a particular time of maximum used as temporal reference. 
For DAV stars, the terms over the second order in $E$ can be neglected, but in PG~1159-035
we need to consider all terms up until the third order in $E$.
Equation~\ref{eq:7} provides a powerful way of determining small $\dot P$ and/or $\ddot P$,
which is known as the {\it (O-C) method}. 
In this method, initial values are first assumed for $T_o$, $P_o$, $\dot P$ and $\ddot P$
   (we used the directly measured $\dot P$ and $\ddot P=0$ as initial values).
Using the initial values and the observed values for $T_{\rm max}$, we then calculate the (integer) 
   number of cycles, $E$, from Eq.~\ref{eq:7}. 
Using the obtained values for $E$, we finally fit Eq.~\ref{eq:7} to the observed $T_{\rm max}$, 
recalculating the values of $T_o$, $P_o$, $\dot P$, and $\ddot P$. 
We note that the number of cycles $E$ calculated in the second step can be biased if the initial
values in the first step are not sufficiently good, since generally they are not. In this case,
we need to take into account the bias calculating the uncertainties in $E$ and considering
all  possible combinations of the numbers of cycle. For each combination, we derive a possible
solution in the  third step. The most likely solution is  the highest probability one
(fitting with lower $\chi^2$).

%%%
Even when  direct measurement cannot be achieved, 
the period versus time plot indicates that the hypothesis stating
that the period  smoothly changes with time is false 
(assuming that the periods and their uncertainties are well determined). 
In this case, we cannot use the (O-C) method.

% ------------------------------------------------------------------------
\section{Period changes in PG~1159-035}   % Sect.3
\label{sect:3} 
% ------------------------------------------------------------------------

%%%
Table~\ref{table:1} lists all pulsation modes detected in 
Fourier Transforms for data of two or more years derived by Costa et al. (2008). 
Dashes ($-$) indicate non-detected modes. 
The periods were calculated at the average BCT date of each data set:
$244\,5346.87422$ for 1983; $244\,6147.66421$ for 1985; $244\,7593.33756$ for 1989;
$244\,9065.92947$ for 1993; and $245\,2410.63535$ for 2002. 
The period uncertainties, $\sigma_P$, were calculated from nonlinear fitting of 
sinusoidal curves for all the detected frequencies, using the Levenberg-Marquardt method. 
Monte Carlo simulations of synthetic light curves 
for PG 1159-035, as proposed by Costa \& Kepler (2000),
indicate that the calculated uncertainties in period (and also in phase 
and amplitude) are excellent when the pulsation frequencies are solved and the amplitudes 
are constant.
However, the calculated uncertainties  can be underestimated 
for modes with amplitudes that change during the observations
or for non-solved frequencies.

% ---------------------------------------------------------------------
\subsection{Direct measurement}  % Sect. 3.1

\begin{figure*}
   \begin{center}
      \begin{tabular}{cccc}
         \resizebox{40mm}{!}{\includegraphics{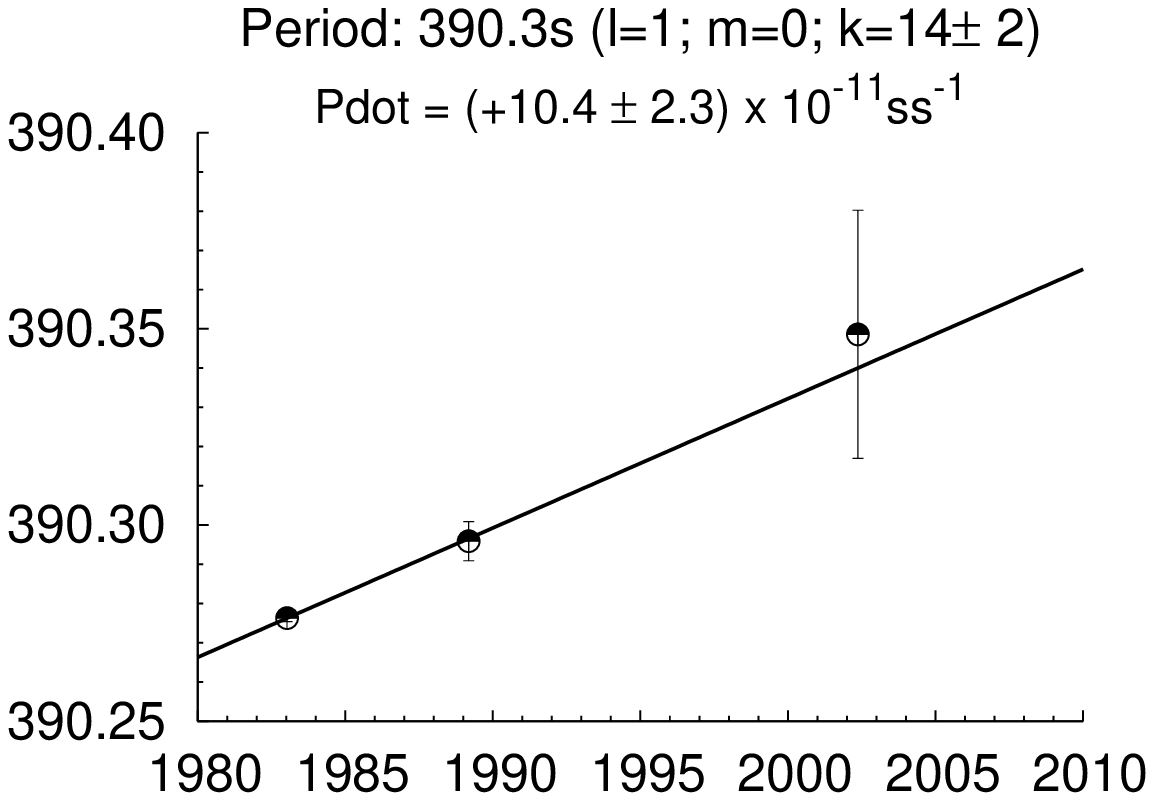}} &
         \resizebox{40mm}{!}{\includegraphics{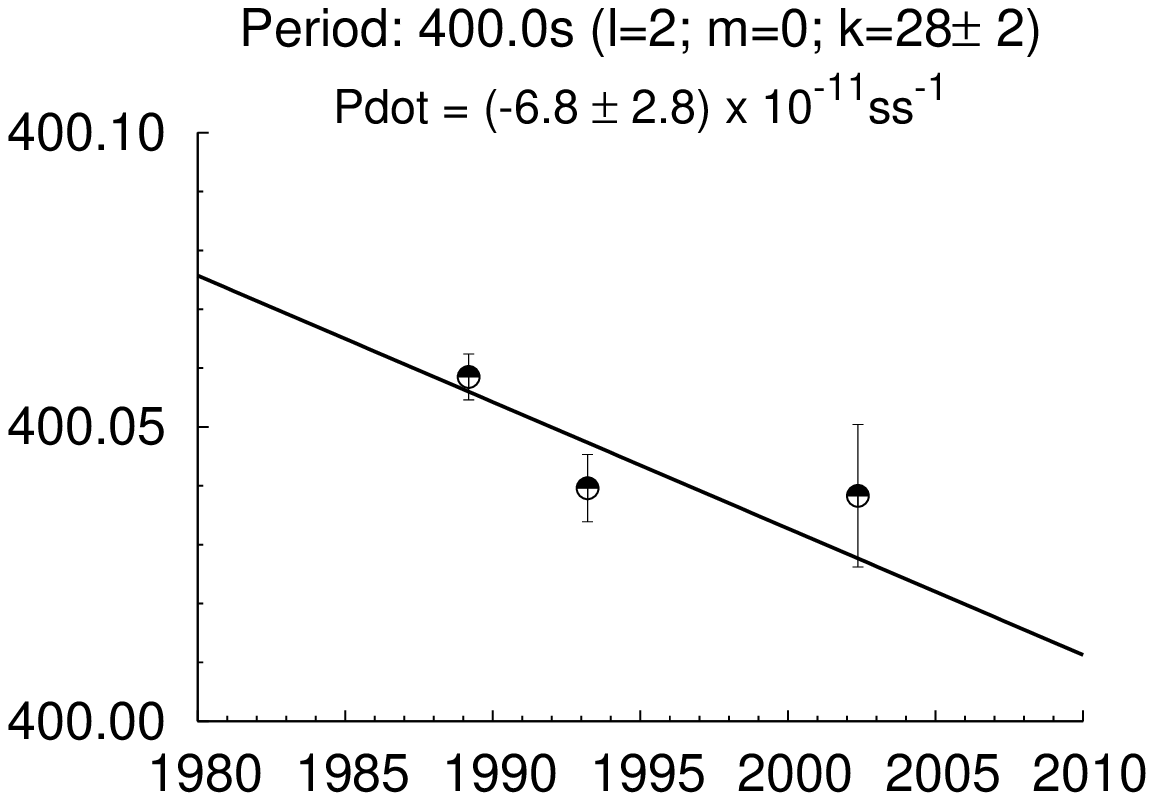}} &
         \resizebox{40mm}{!}{\includegraphics{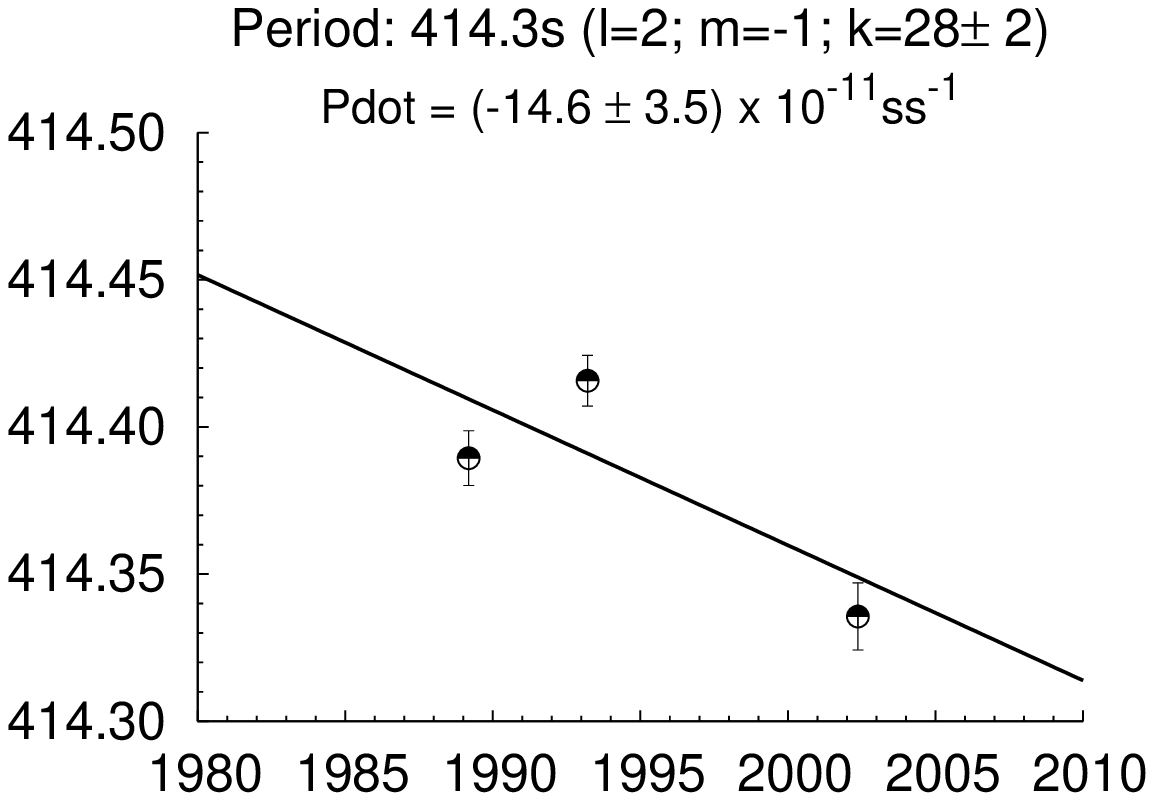}} &
         \resizebox{40mm}{!}{\includegraphics{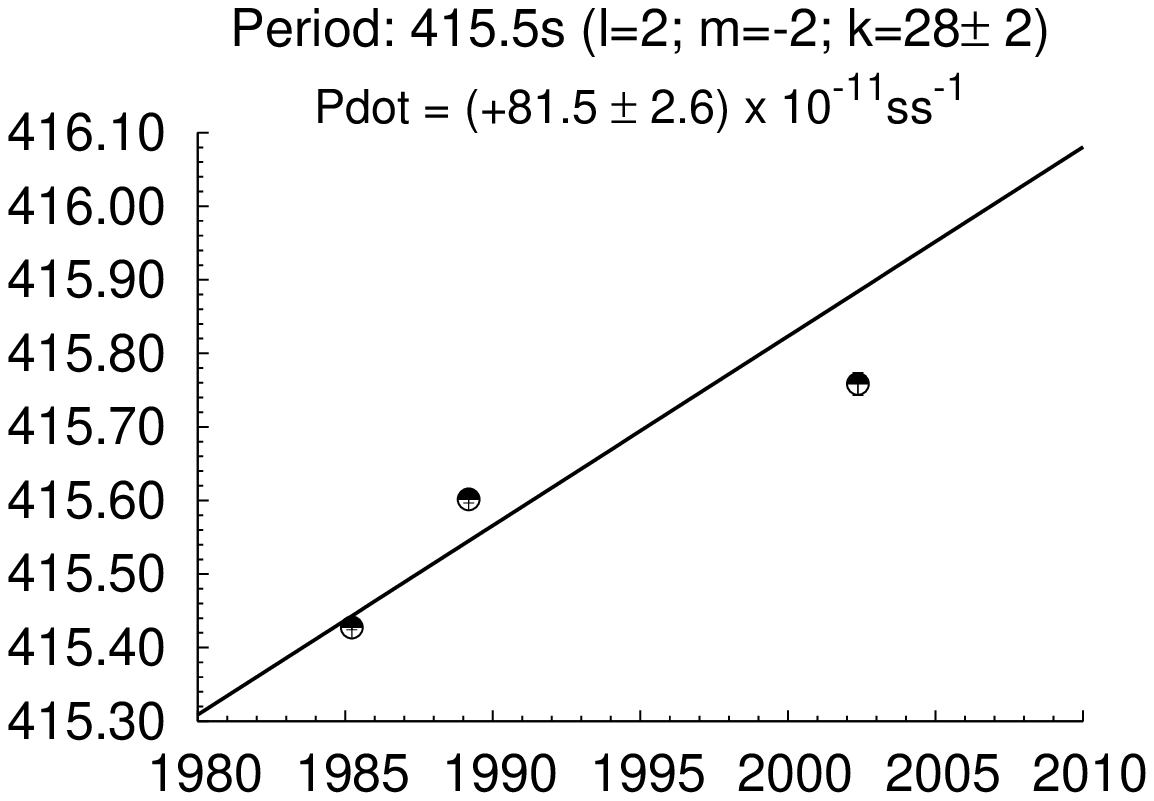}} \\
         \resizebox{40mm}{!}{\includegraphics{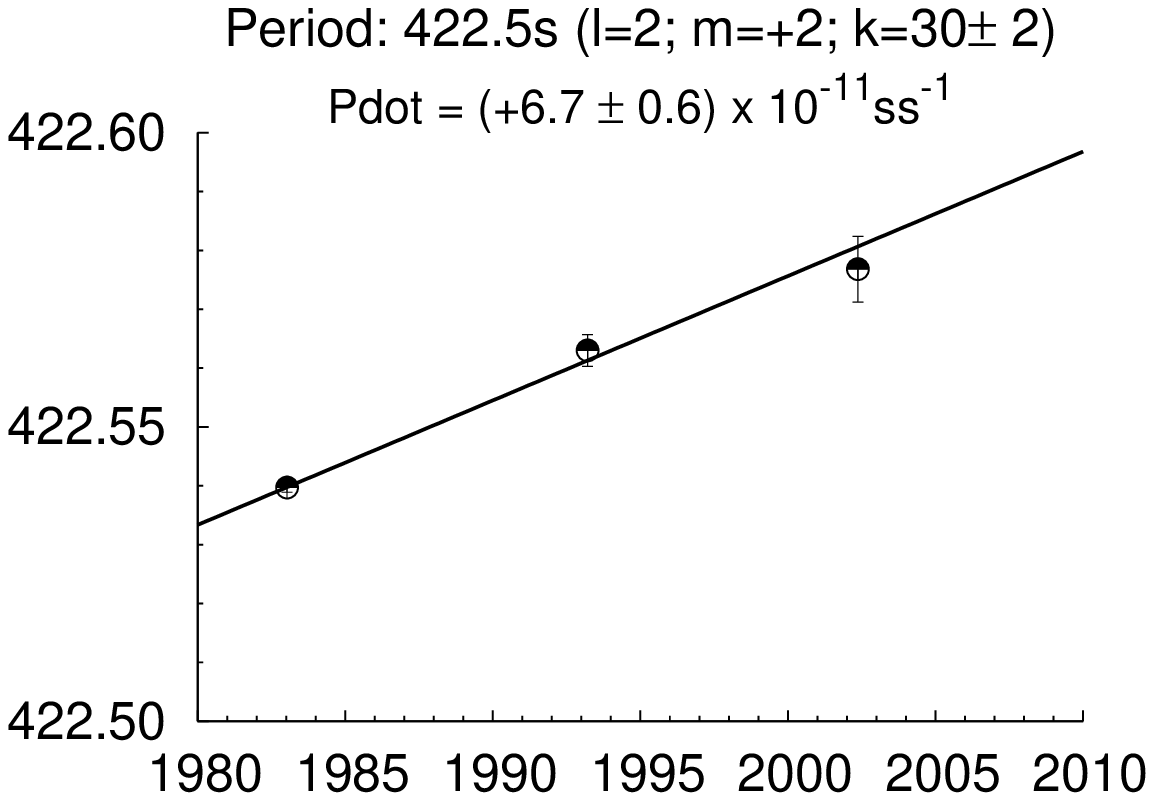}} &
         \resizebox{40mm}{!}{\includegraphics{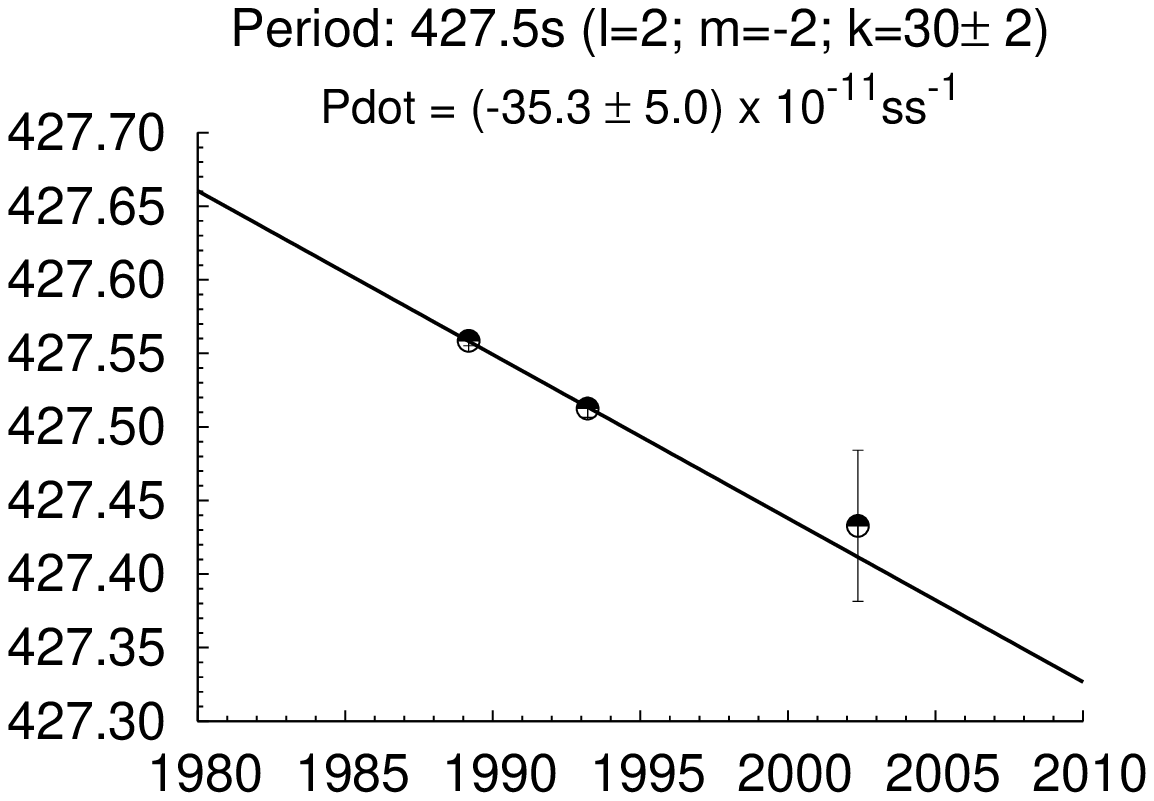}} &
         \resizebox{40mm}{!}{\includegraphics{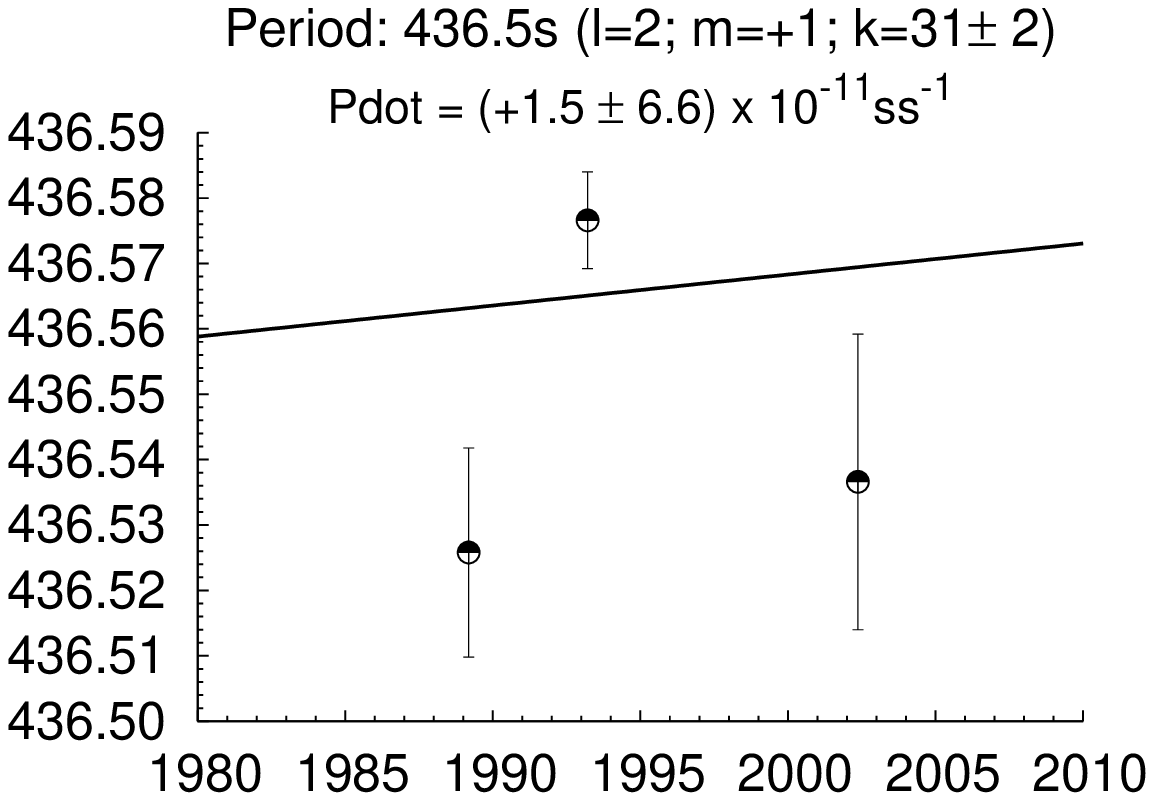}} &
         \resizebox{40mm}{!}{\includegraphics{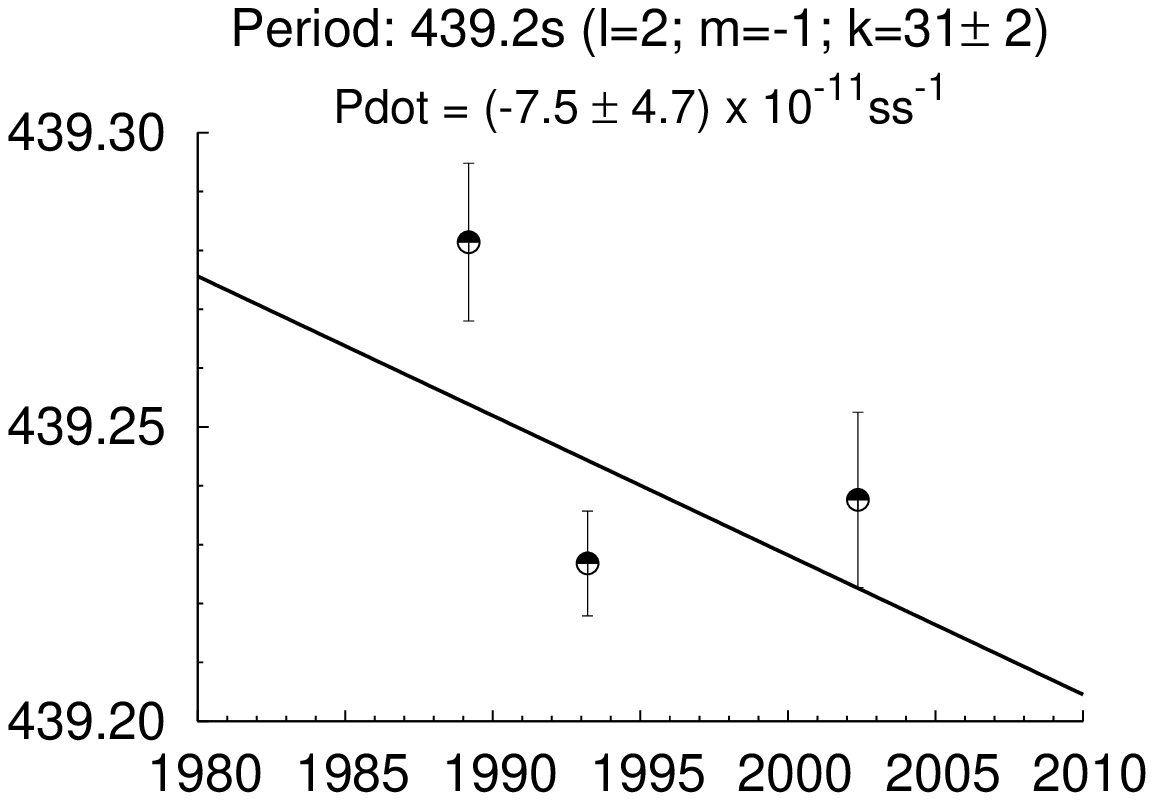}} \\
         \resizebox{40mm}{!}{\includegraphics{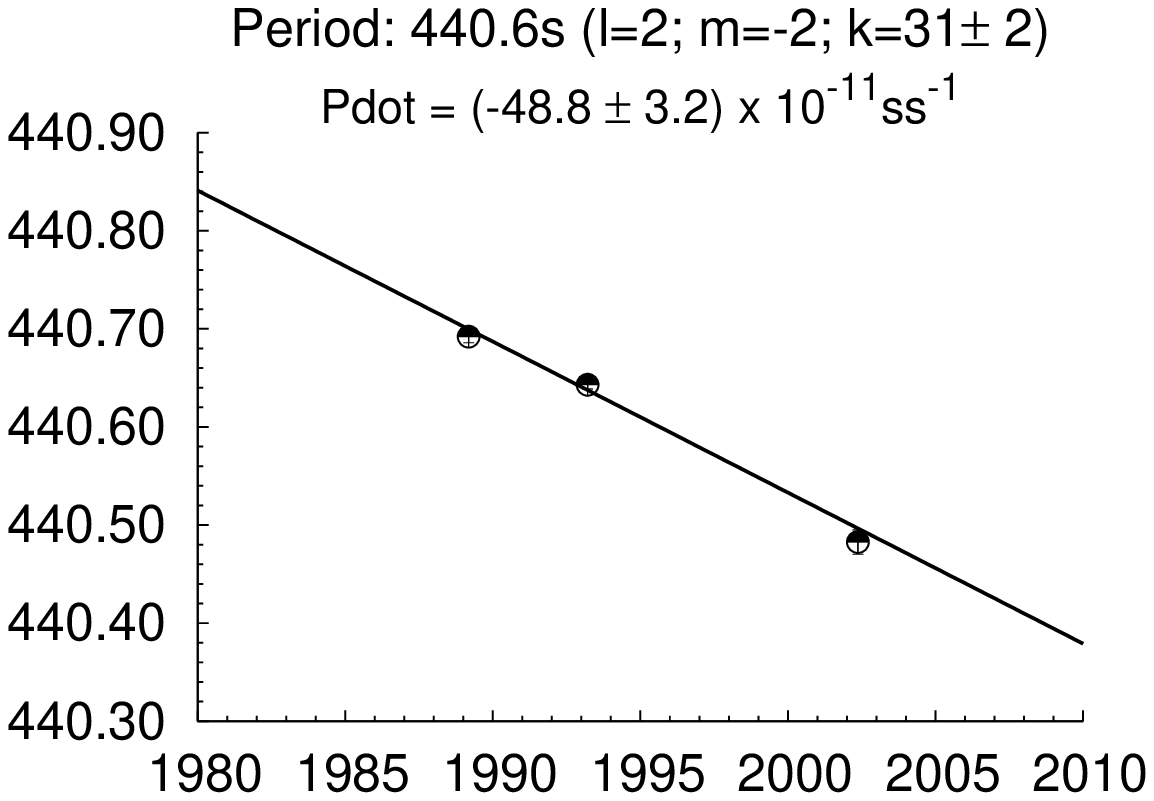}} &
         \resizebox{40mm}{!}{\includegraphics{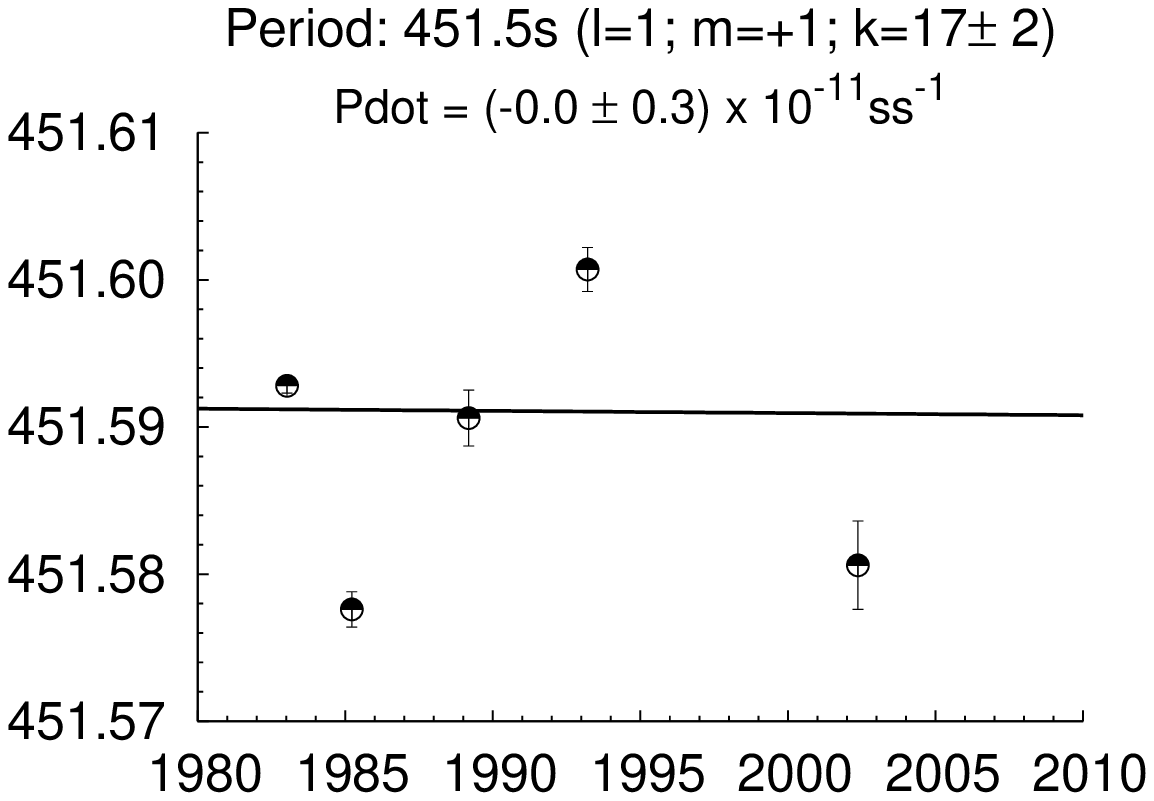}} &
         \resizebox{40mm}{!}{\includegraphics{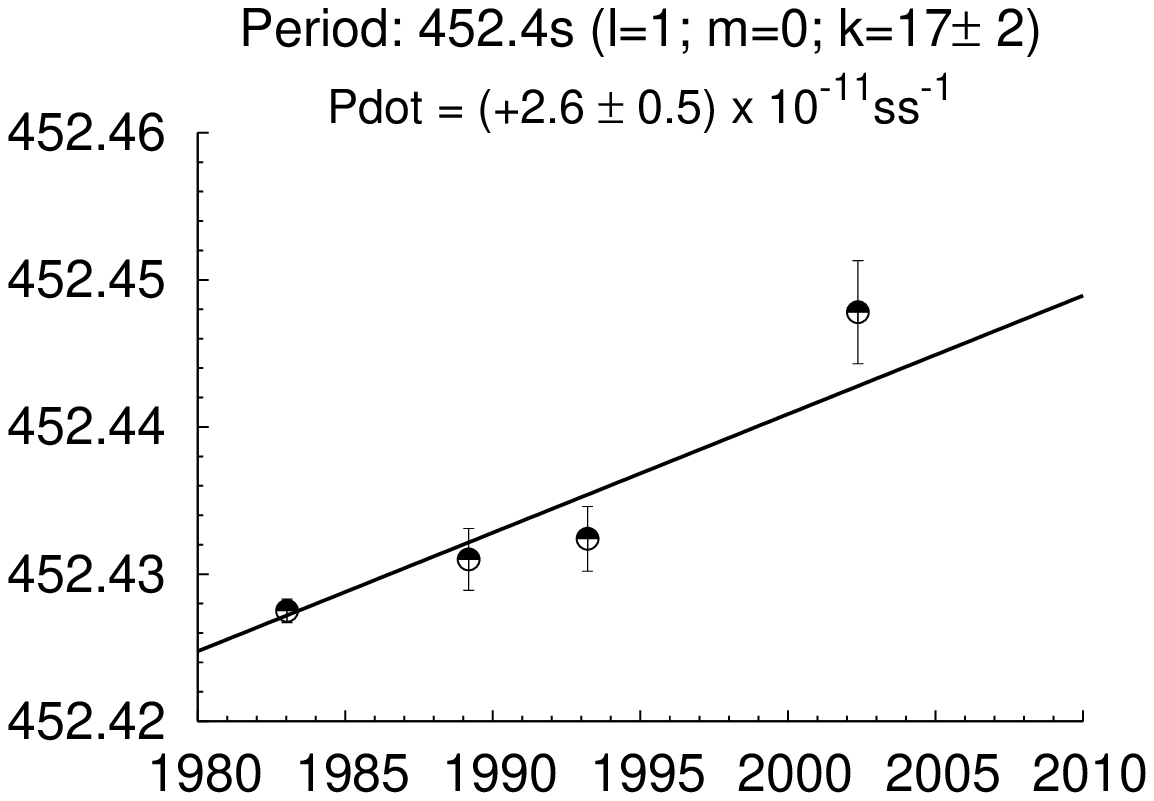}} &
         \resizebox{40mm}{!}{\includegraphics{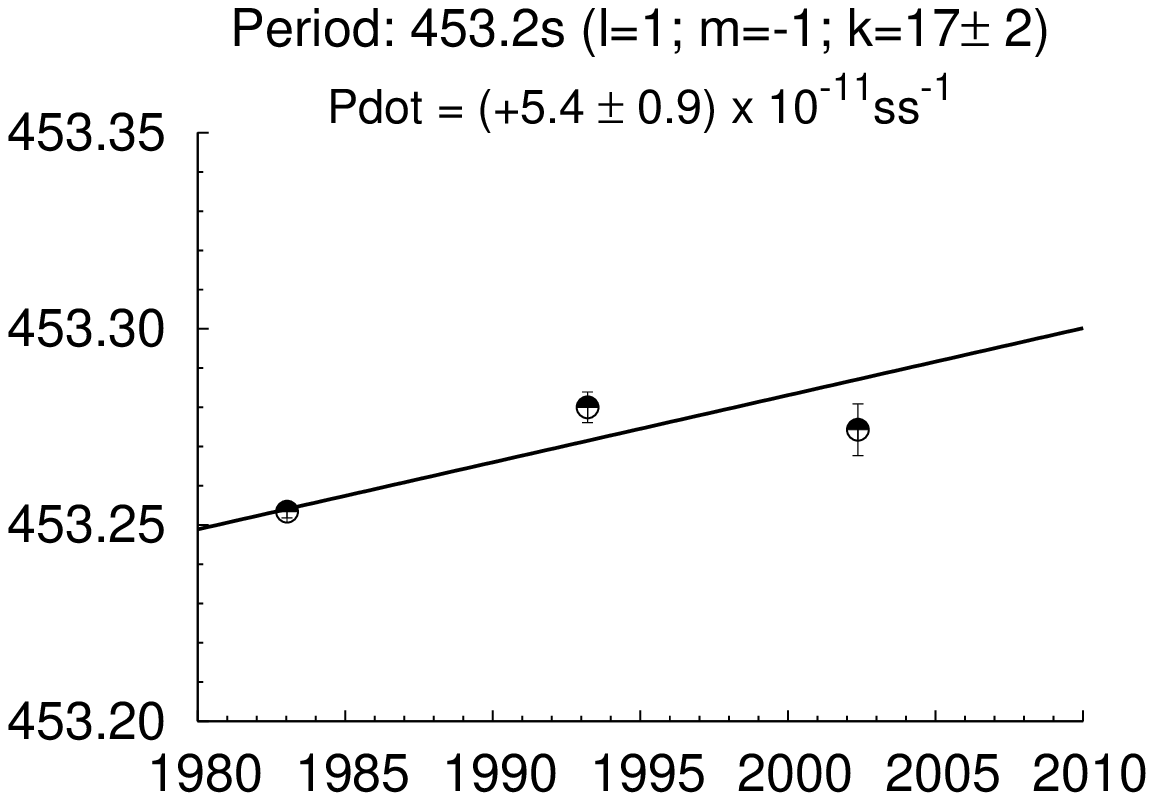}} \\
         \resizebox{40mm}{!}{\includegraphics{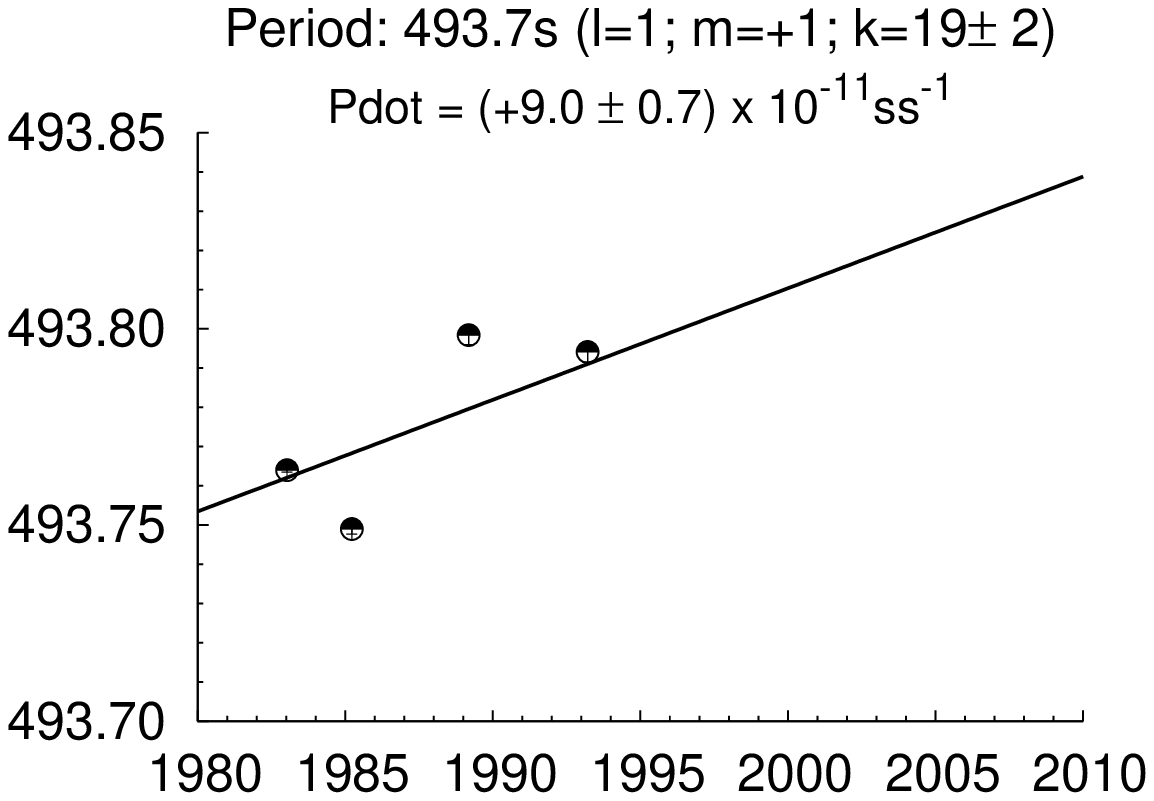}} &
         \resizebox{40mm}{!}{\includegraphics{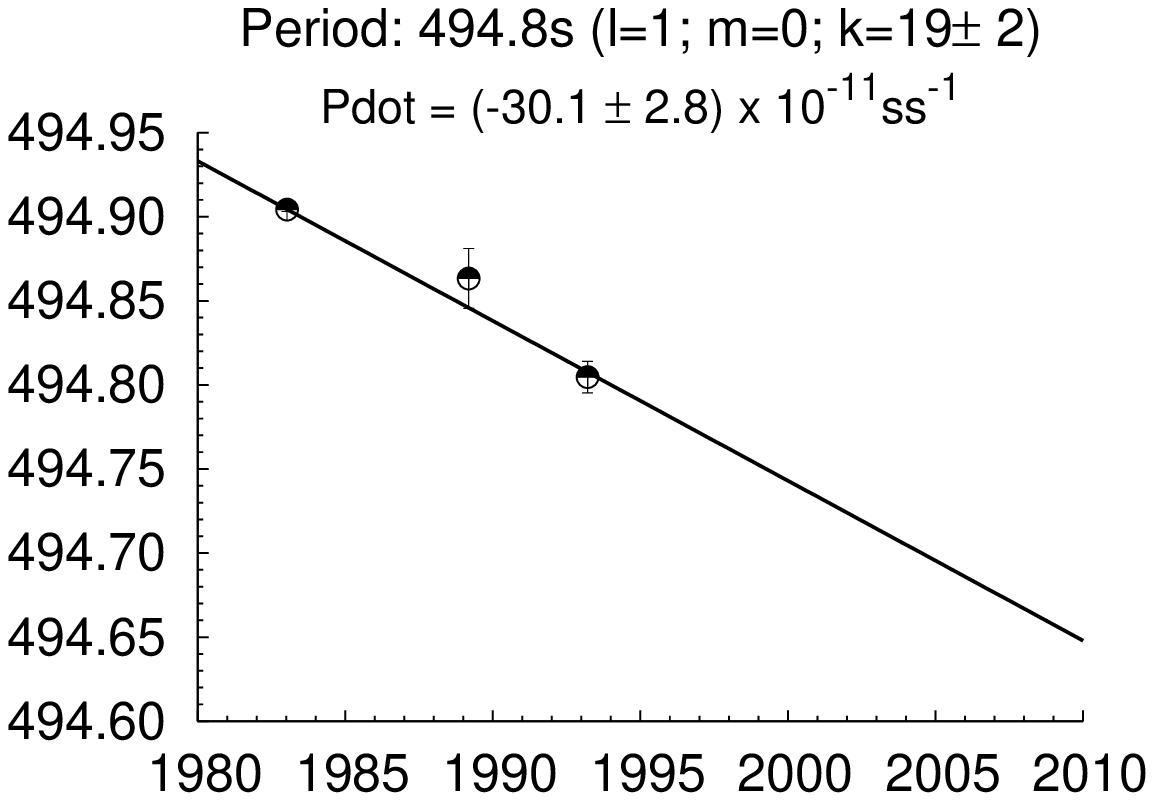}} &
         \resizebox{40mm}{!}{\includegraphics{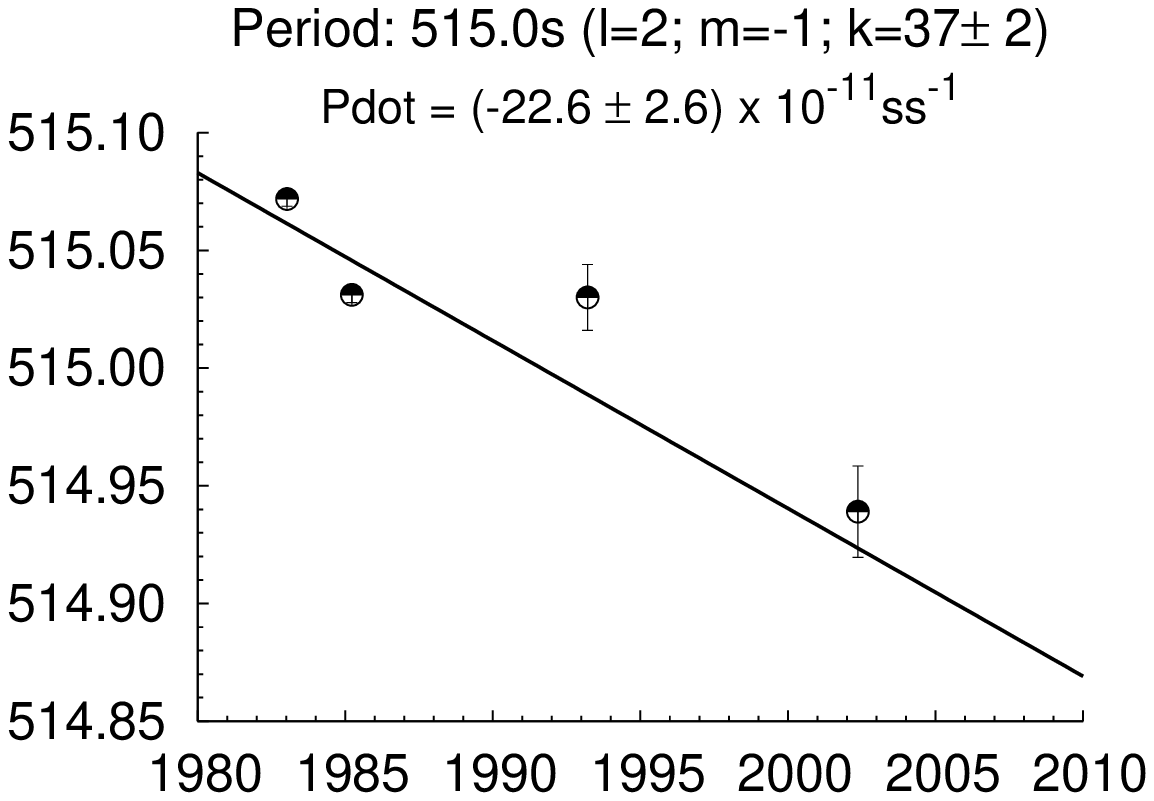}} &
         \resizebox{40mm}{!}{\includegraphics{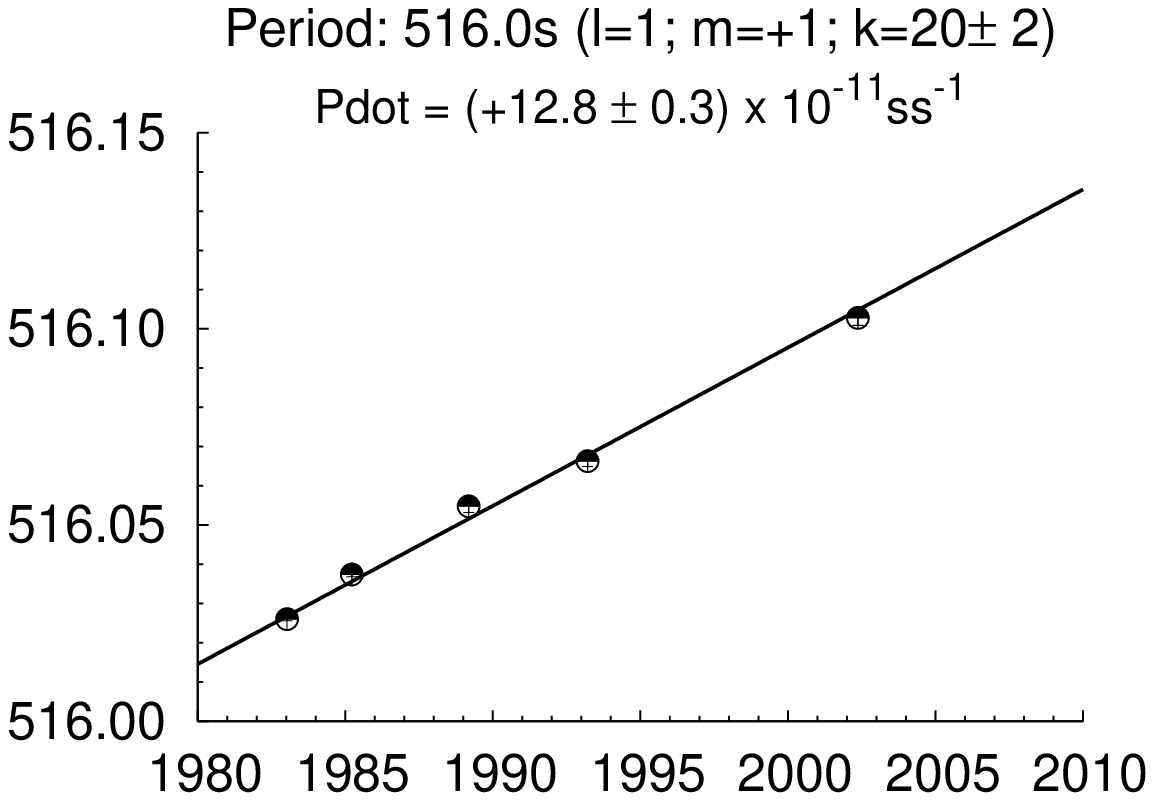}} \\
         \resizebox{40mm}{!}{\includegraphics{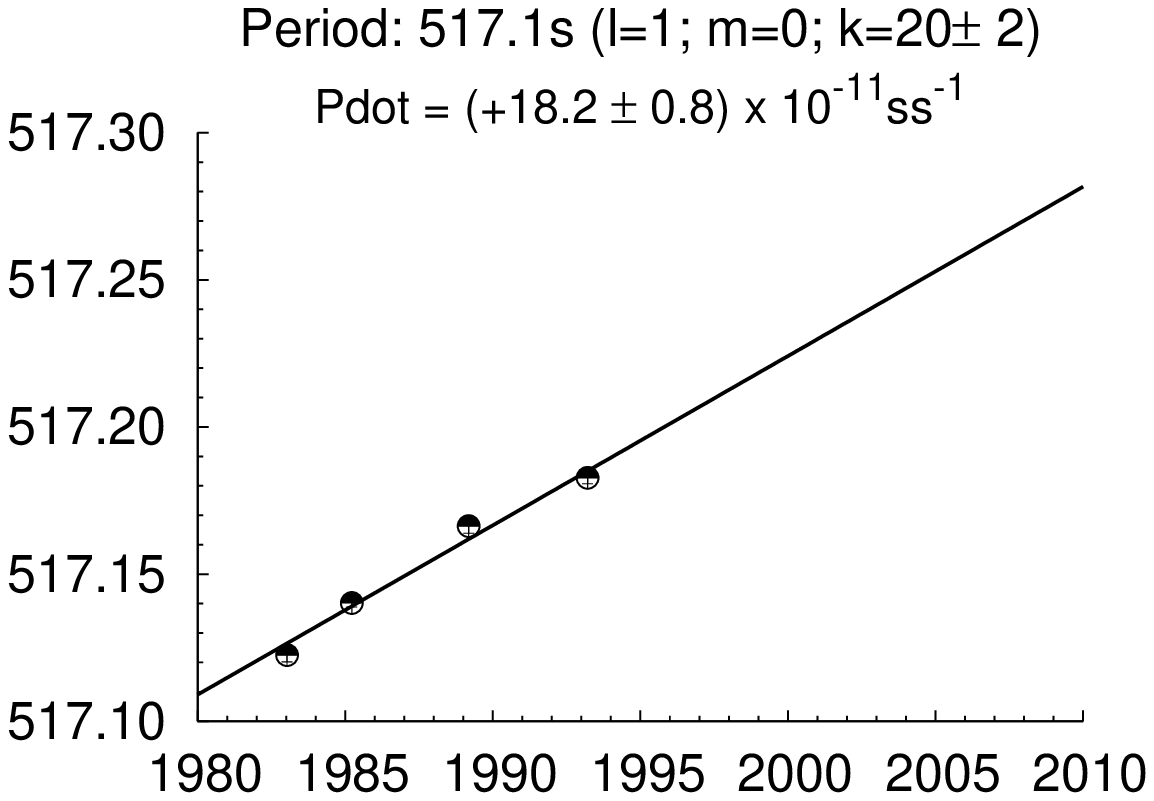}} &
         \resizebox{40mm}{!}{\includegraphics{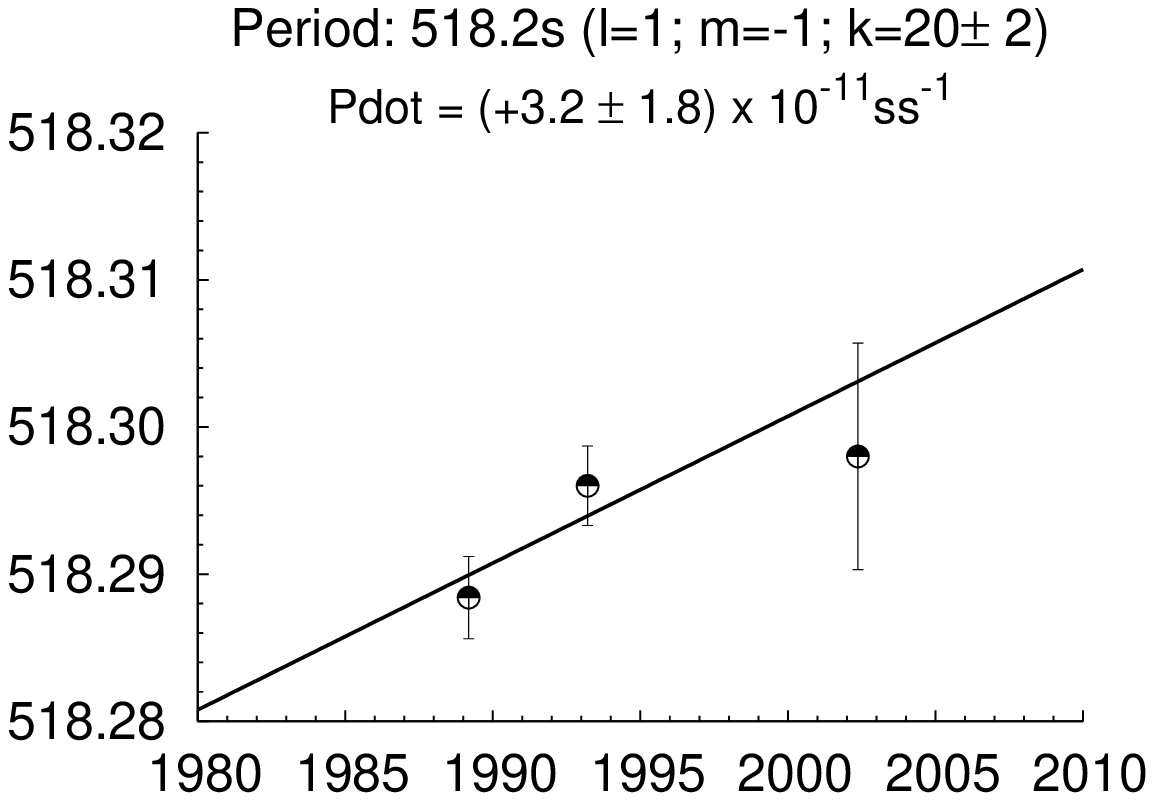}} &
         \resizebox{40mm}{!}{\includegraphics{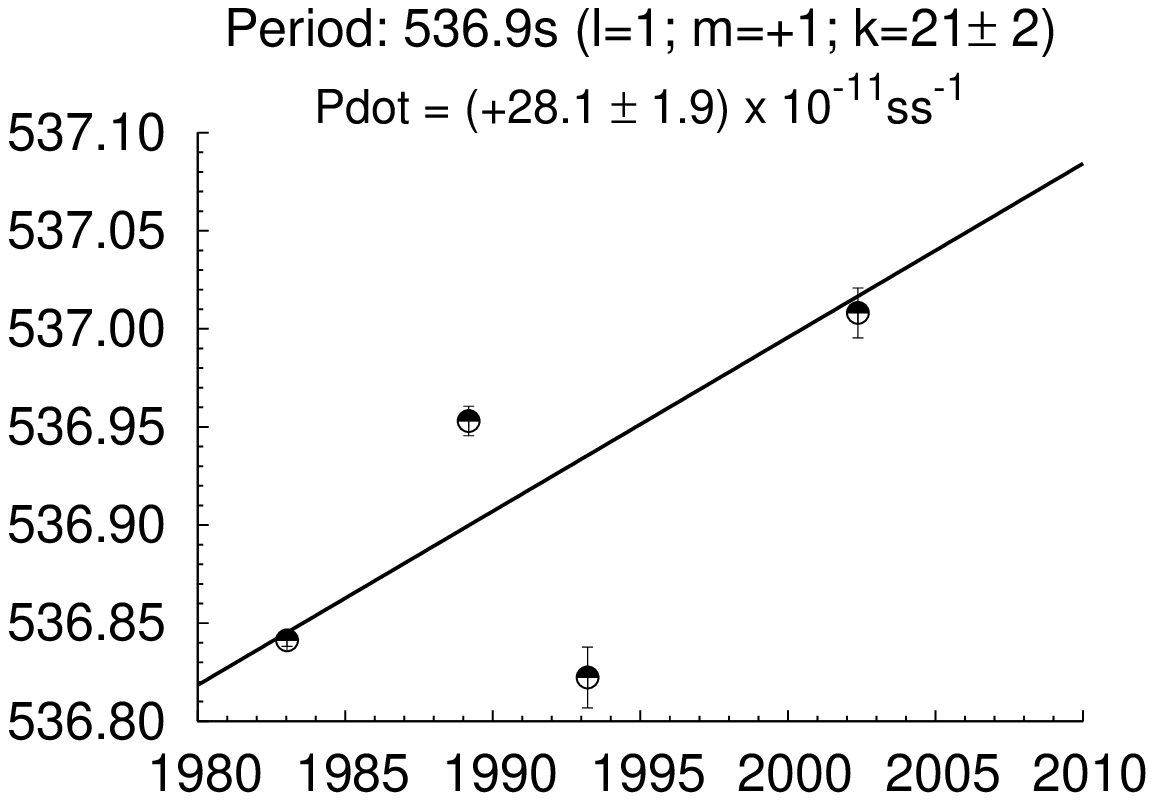}} &
         \resizebox{40mm}{!}{\includegraphics{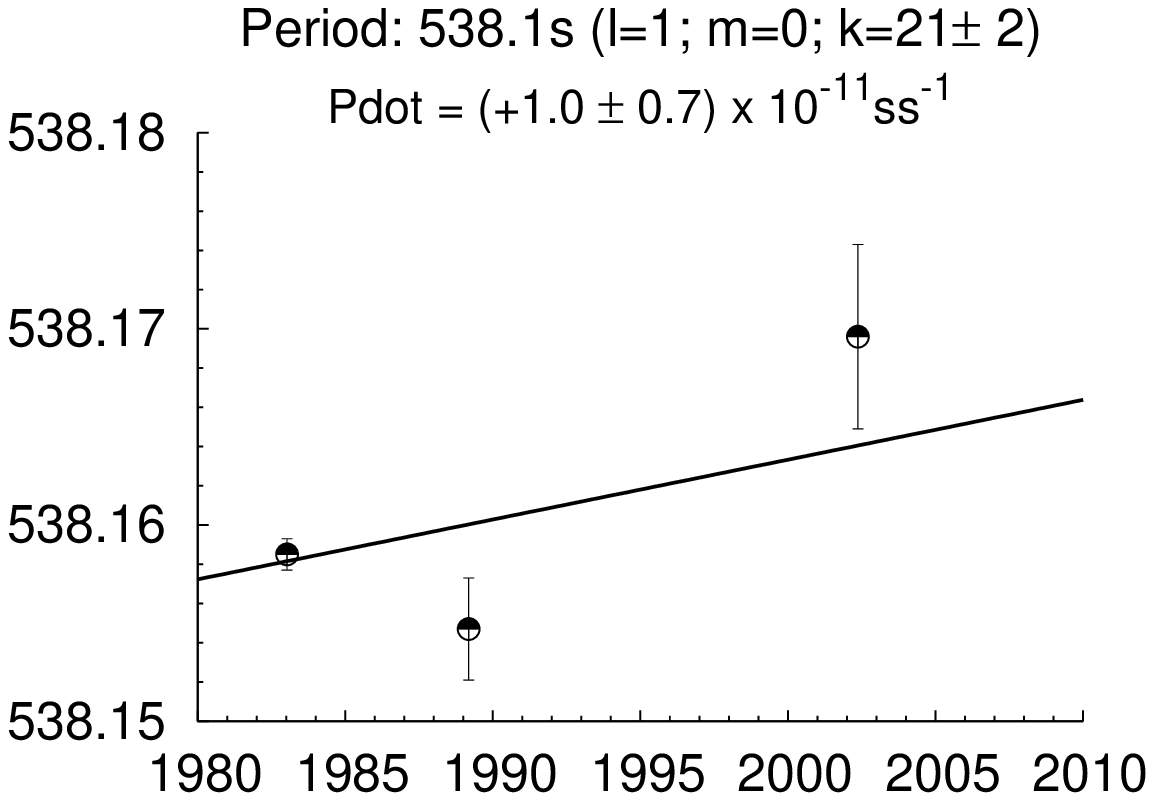}} \\
         \resizebox{40mm}{!}{\includegraphics{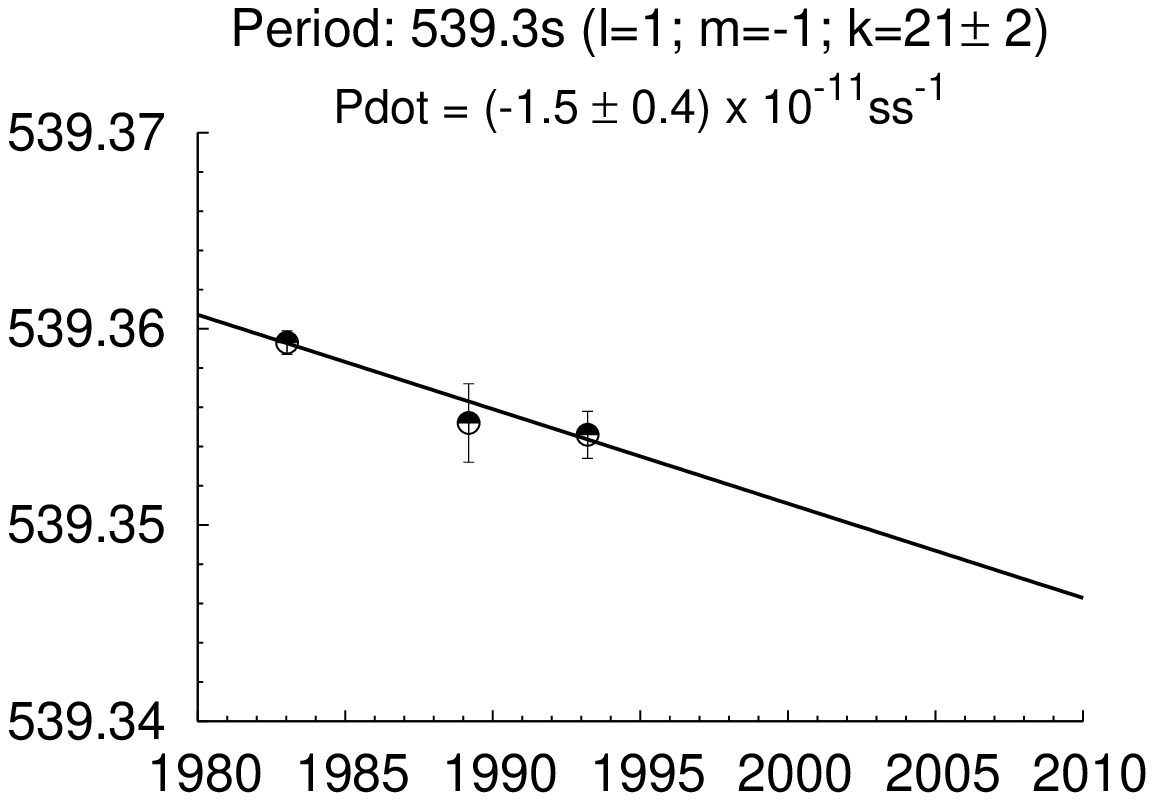}} &
         \resizebox{40mm}{!}{\includegraphics{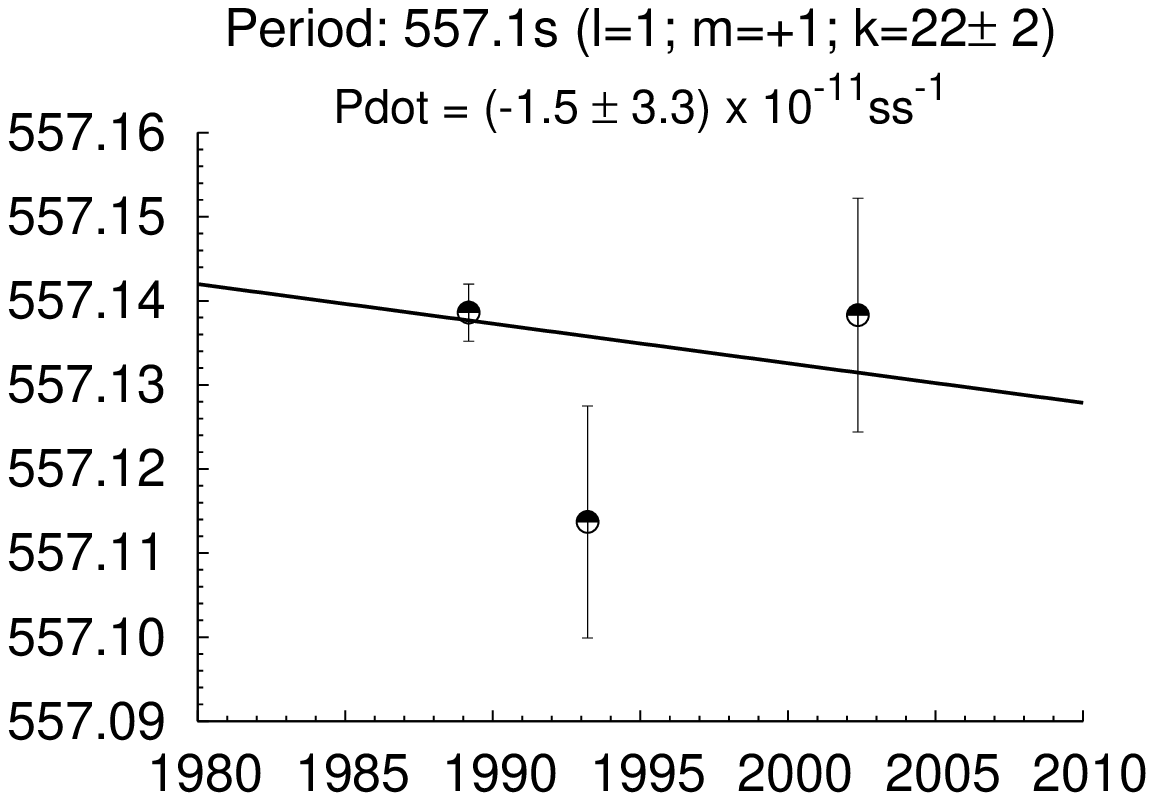}} &
         \resizebox{40mm}{!}{\includegraphics{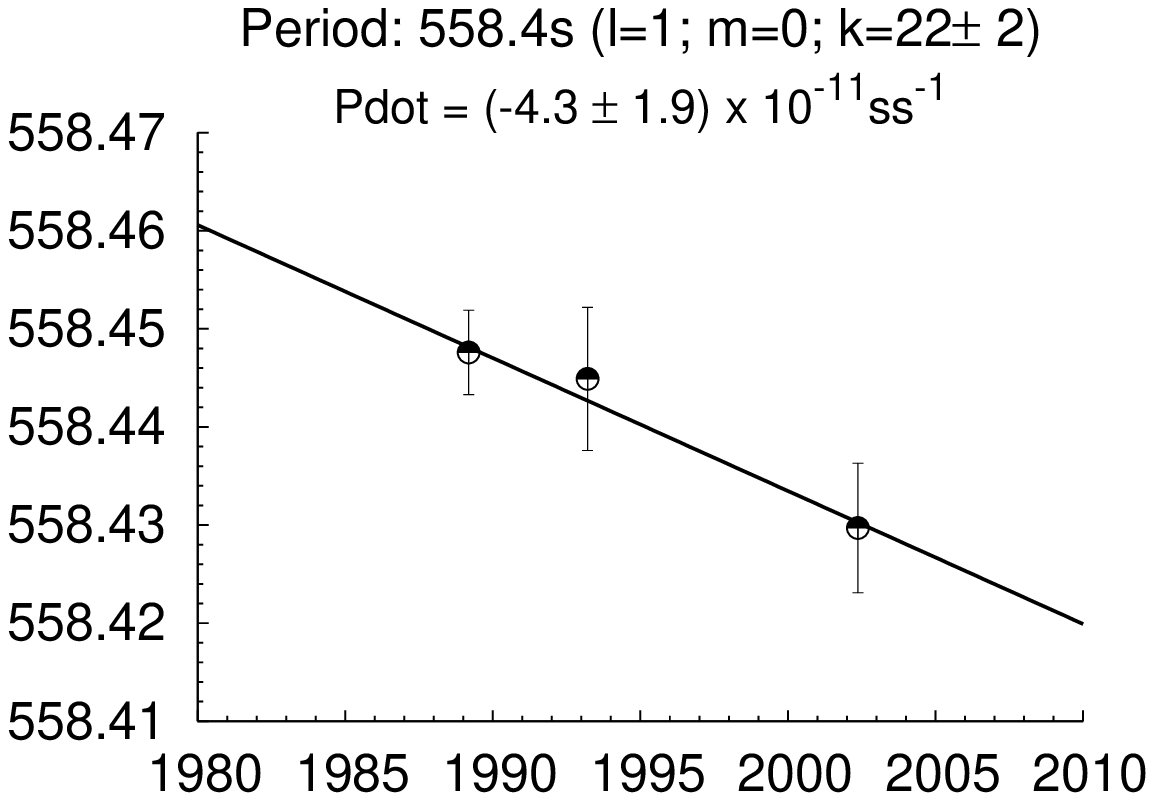}} &
         \resizebox{40mm}{!}{\includegraphics{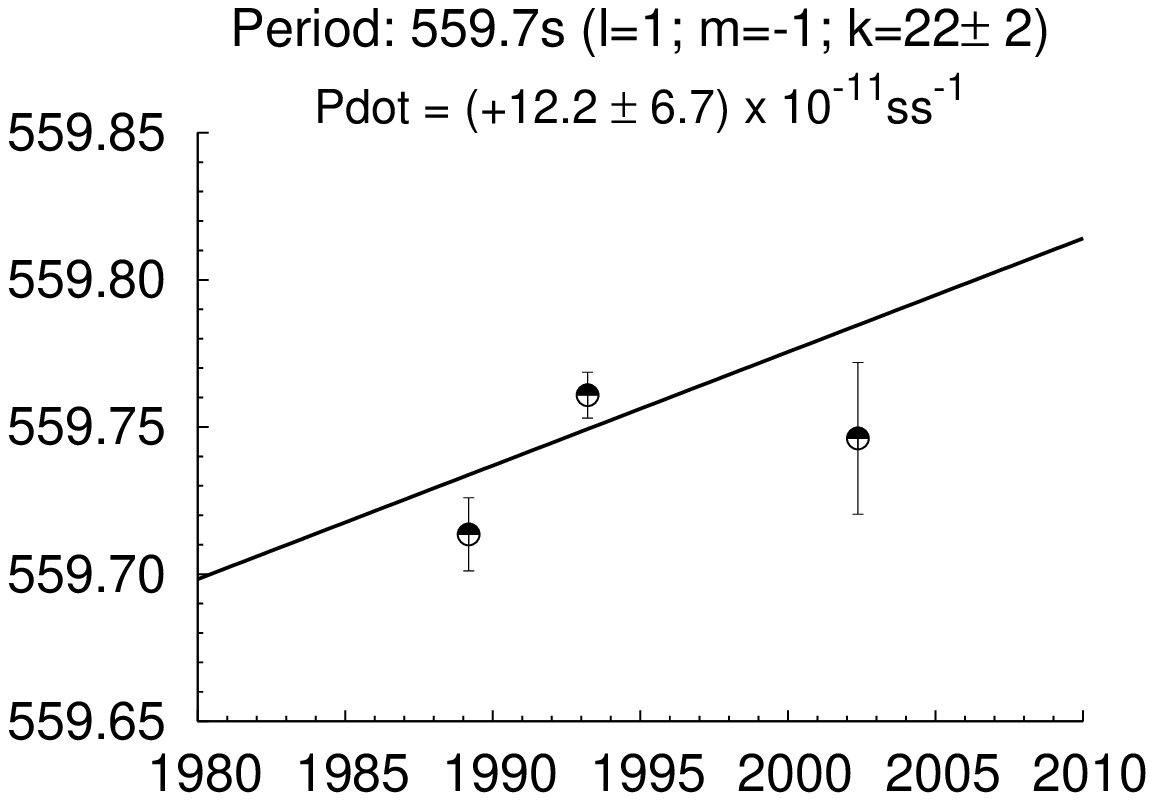}} \\
         \resizebox{40mm}{!}{\includegraphics{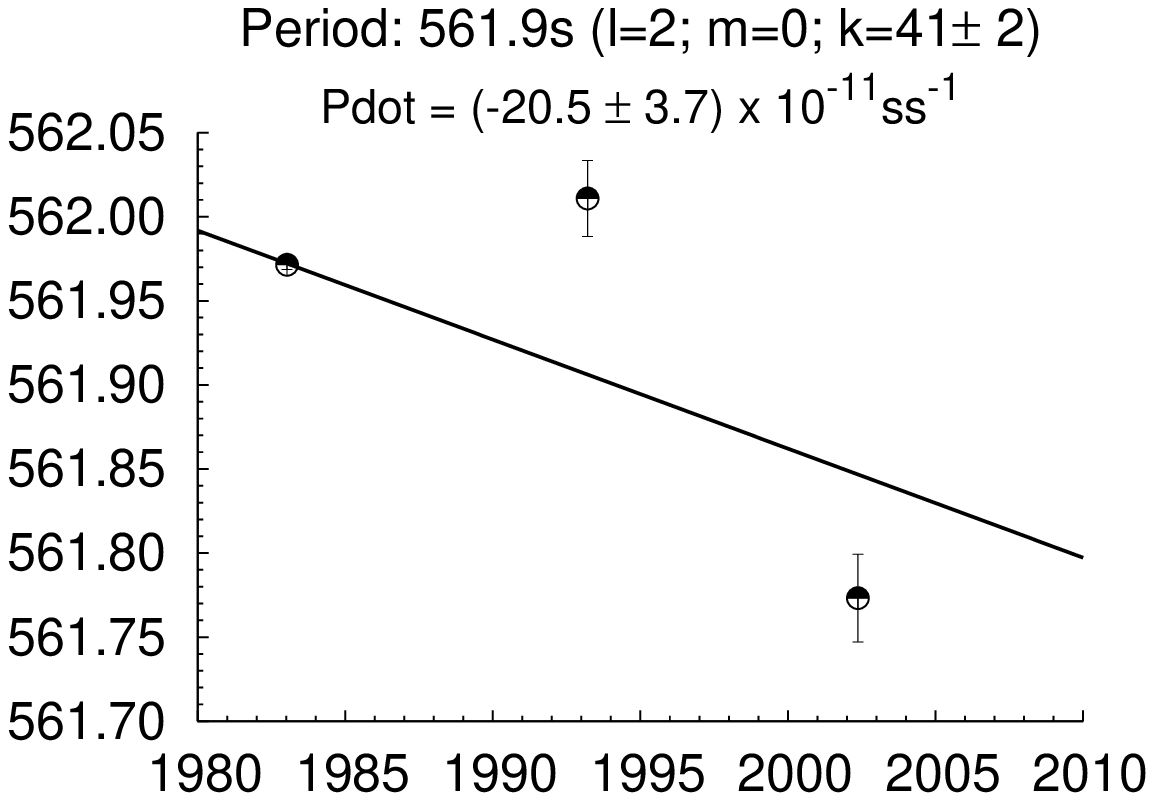}} &
         \resizebox{40mm}{!}{\includegraphics{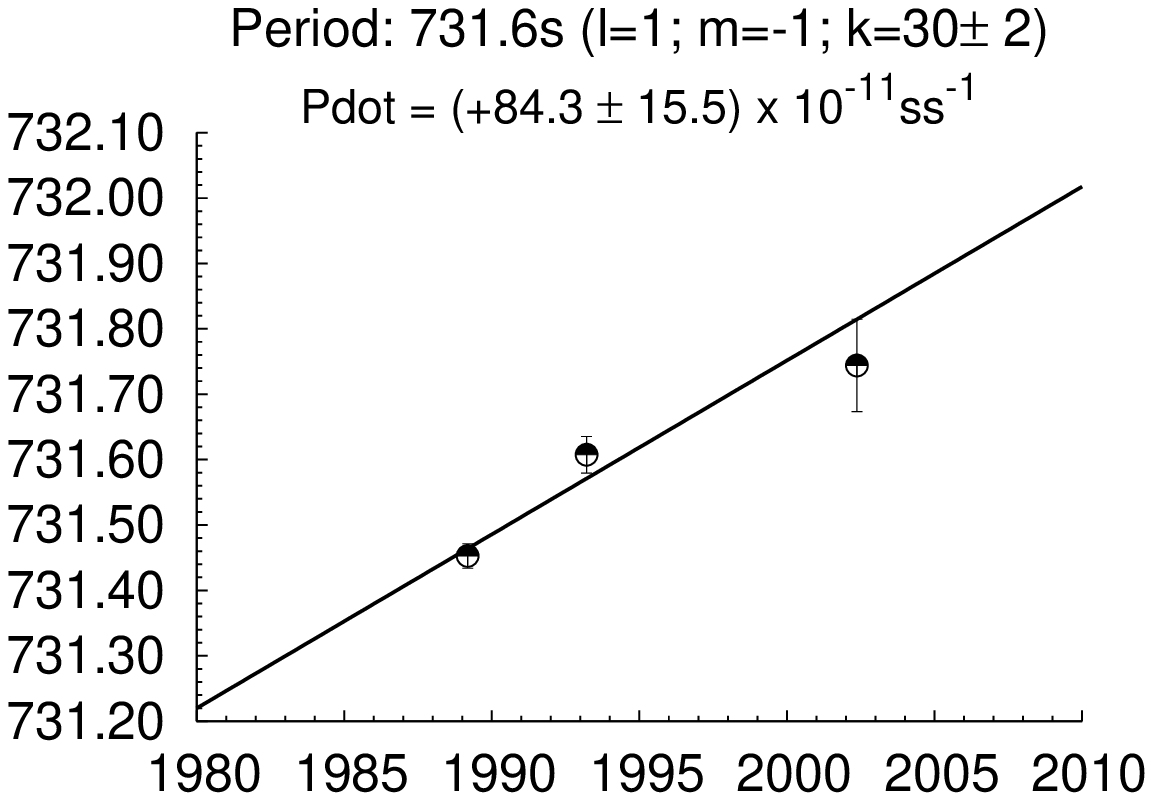}} &
         \resizebox{40mm}{!}{\includegraphics{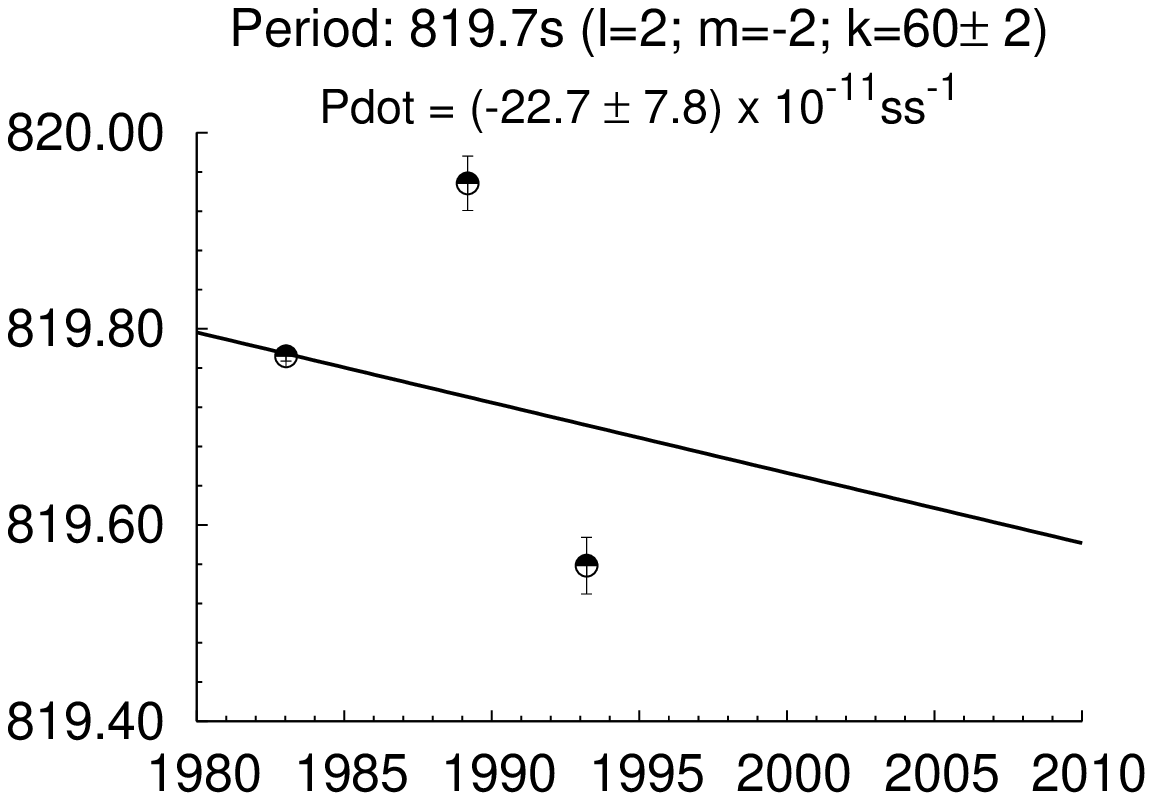}} &
      \end{tabular}
      \caption{Temporal changing in the periods of 27 PG~1159-035 pulsation modes. 
               The vertical axis of each graph represents the periods, in seconds, 
               and the horizontal axis is the time,  in years. The bars represent
	       the $\pm 1\sigma$ uncertainties. The mode identification 
               (period in seconds and the $\ell$, $m$ and $k$ indices) appears in 
               the first line above each graph. In the second line ``Pdot'' 
               ($\dot P$) is the slope of the straight line fitted to the points.}
      \label{fig:2}
   \end{center}
\end{figure*}

%%%
Figure~\ref{fig:2} shows the period (in seconds) 
versus time (in years) plots for all the modes detected in three or more FTs. 
Above each graph, there is the average period (in seconds) and the ($\ell$, $m$, $k$) indices
of the pulsation mode (the $k$ index has an uncertainty $\pm 2$). 
The error bars represent the  $1\sigma$ uncertainties in the period measurement. 
The slope of the fitted curve is indicated at the top of each graph.

%%%
In  most cases, the points distribution is consistent statistically with a linear fit, 
taking into account the uncertainties in the period measurements. 
This is particularly notable in the cases of the 422.5 s, 427.5 s, 440.6 s, 494.8 s, 516.0 s, and 517.1 s
pulsation modes. In the 451.5 s, 493.7 s, 536.9 s, and 819.7 s plots, the points are scattered 
with large dispersions relative to the straight lines. 
It is possible that some  points with lower confidence levels have been mistakenly identified.
The measured $\dot P$ for the pulsation modes in Fig.~\ref{fig:2}
are listed in Table~\ref{table:2}. 
The relative standard uncertainty ($\sigma_{\dot P}/\,|{\dot P}\,|$) in the $\dot P$ measurement of 
the 516.0 s period is only 2.0\% and for other five periods the relative standard 
uncertainties are less than $10\%$: 422.5 s, 440.6 s, 452.4 s, 453.2 s, and 517.1 s.

% ------------------------------------------------------------------------
\subsection{(O-C) method} % Sect. 3.2

%%%
We used the directly measured $\dot P$ as initial values (setting $\ddot P =0$) in the (O-C) method to refine 
the $\dot P$ and  periods of the modes present in three or more FTs. 
With three points only, the fitting  provides preliminary values  for future calculations.  
We also used the (O-C) method to calculate $\ddot P$ of the modes
present in four or five of the FTs.
Our most accurate results are given in Table~\ref{table:2}.

%%%
The accuracy of the $\dot P$ determination is, in general, 
two orders of magnitude smaller than that of direct measurement. 
We note that the (O-C) fitting assumes that the periods change smoothly with time. 
If the changes are not smooth or if they depend on  higher order terms, 
the (O-C) can produce incorrect results. 
It is also important to note that although the most suitable
solution is the most probable one, it is not necessarily the true solution
and no solution has an uncertainty of below 1\%.
Future photometric observations of PG~1159-035 should confirm or discard our results.

%%%
We comment on three particular cases: changes in the 516.0 s, 517.1 s, and 539.1 s periods. 
The two first cases were identified by W91 and Costa et al. (2008) 
as the $m=+1$ and $m=0$ components, respectively, of the  triplet ($\ell=1$) 
with radial index $k=20\pm2$; they are used in Sect.~\ref{sect:6} 
to calculate PG~1159-035 evolutionary rates. 
The 516.0 s mode is clearly present in the FT of all data sets.
The same occurs with the 517.1 s mode, apart from the 2002 FT. 
The 539.1 s period was identified as  the component $m=-1$ of
the $k=21\pm 2$ triplet and was the second period to have its $\dot P$ determined (Costa et al. 1995).

% FIG. 1 ---------------------------------------
\begin{figure}
\centerline{\epsfig{figure={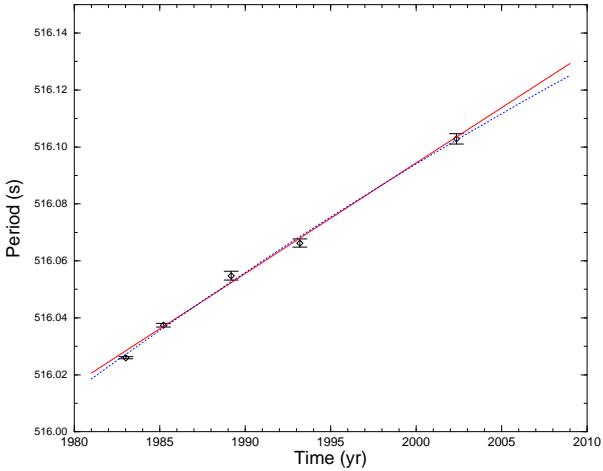},width=8.0cm}}
\caption{Temporal change in the 516.0 s pulsation period. 
         The solid line represents a linear fitting and the dashed line 
         a quadratic fitting.} 
\label{fig:1} 
\end{figure}
% ----------------------------------------------

% FIG. 2  -----------------------------------------------------------------------------
% (53mm is ok for 3 cols)

% TABLE 2 ------------------------------------
\begin{table*}
\caption{Results from the direct method and (O-C) fitting.} 
\label{table:2}
{\scriptsize
 \begin{center}
   \begin{tabular}{ccrc|c|r|ccrr} \hline\hline 
          &        &     &     &     &                      &         &       &                                 &  \\
    \multicolumn{4}{c|}{Pulsation Mode} &
    \multicolumn{1}{c}{} &
    \multicolumn{1}{|c|}{Direct} &
    \multicolumn{4}{c}{(O-C) Fitting} \\ \cline{1-4} \cline{6-10}
          &        &     &     &     &                      &         &       &                                 &  \\
    \multicolumn{1}{c}{$\langle P\rangle$}     & \multicolumn{1}{c}{$\ell$}   & \multicolumn{1}{c}{$m$} & \multicolumn{1}{c}{$k$} &
    \multicolumn{1}{|c|}{$N$}     & \multicolumn{1}{c|}{$\dot P$}& 
    \multicolumn{1}{c}{ $T_o$}    & \multicolumn{1}{c}{$P_o$}    & \multicolumn{1}{c}{$\dot P$} & \multicolumn{1}{c}{$\ddot P$} \\

    \multicolumn{1}{c}{(s)}   & \multicolumn{2}{c}{}    &  \multicolumn{1}{c}{$\pm 2$} &
    \multicolumn{1}{|c|}{} & \multicolumn{1}{c|}{$\times 10^{-11}\,{\rm ss}^{-1}$} &
    \multicolumn{1}{c}{(BCT)} & \multicolumn{1}{c}{(s)} &  \multicolumn{1}{c}{$\times 10^{-11}\,{\rm ss}^{-1}$} &
    \multicolumn{1}{c}{ $\times 10^{-19}\,{\rm ss}^{-2}$} \\ \hline
          &        &     &     &     &                      &         &       &                                 &  \\
    % Pdots calculados com o_c.tcl
% P      Ndata      Tmax_o  (BCT)  Po   (sec) Pdot (x 1.e-11 s/s)                  dP2/dt2  (x1.e-19 s/s**2)
390.3 & 1 & 0 & 14 & 3 & $+10.4\pm 2.3$ & $2447593.33753\pm 0.00009$ & $390.30088\pm 0.00003$ & $  12.596\pm 0.020$ &  \\
400.0 & 2 & 0 & 28 & 3 & $ -6.8\pm 2.8$ & $2449065.92791\pm 0.00012$ & $400.04271\pm 0.00002$ & $  -0.123\pm 0.027$ &  \\
414.3 & 2 &-1 & 28 & 3 & $-14.6\pm 3.5$ & $2449065.92697\pm 0.00018$ & $414.41413\pm 0.00004$ & $ -22.065\pm 0.042$ &  \\    % ruim
415.5 & 2 &-2 & 28 & 3 & $+81.5\pm 2.6$ & $2447593.33537\pm 0.00008$ & $415.60748\pm 0.00008$ & $  55.633\pm 0.042$ &  \\    % ruim
422.5 & 2 &+2 & 30 & 3 & $ +6.7\pm 0.6$ & $2449065.92752\pm 0.00006$ & $422.56237\pm 0.00001$ & $   4.984\pm 0.008$ &  \\     %OK
427.5 & 2 &-2 & 30 & 3 & $-35.3\pm 5.0$ & $2449065.93108\pm 0.00012$ & $427.51584\pm 0.00004$ & $ -33.956\pm 0.060$ &  \\    %OK?
436.5 & 2 &+1 & 31 & 3 & $ +1.5\pm 6.6$ & $2449065.92902\pm 0.00015$ & $436.59526\pm 0.00006$ & $ -10.899\pm 0.052$ &  \\    
439.2 & 2 &-1 & 31 & 3 & $ -7.5\pm 4.7$ & $2449065.92774\pm 0.00017$ & $439.23003\pm 0.00005$ & $  -9.066\pm 0.048$ &  \\
440.6 & 2 &-2 & 31 & 3 & $-48.8\pm 3.2$ & $2449065.92750\pm 0.00008$ & $440.64399\pm 0.00002$ & $ -47.992\pm 0.025$ &  \\
%451.5 & 1 &+1 & 17 & 5 & $  0.0\pm 0.3$ & $2449065.92845\pm 0.00002$ & $451.60064\pm 0.00000$ & $  -1.715\pm 0.005$ &  \\
451.5 & 1 &+1 & 17 & 5 & $  0.0\pm 0.3$ & $2449065.92845\pm 0.00003$ & $451.60156\pm 0.00001$ & $  -1.498\pm 0.005$ & $-2.908\pm 0.001$ \\
452.4 & 1 & 0 & 17 & 4 & $ +2.6\pm 0.5$ & $2449065.93009\pm 0.00004$ & $452.43494\pm 0.00002$ & $   4.720\pm 0.007$ & $  -0.333\pm 0.014$ \\
453.2 & 1 &-1 & 17 & 3 & $ +5.4\pm 0.9$ & $2449065.92972\pm 0.00007$ & $453.28156\pm 0.00002$ & $   4.143\pm 0.013$ &  \\
493.7 & 1 &+1 & 19 & 4 & $ +9.0\pm 0.7$ & $2447593.33963\pm 0.00003$ & $493.79587\pm 0.00007$ & $  15.102\pm 0.041$ & $8.656\pm 0.173$ \\
494.8 & 1 & 0 & 19 & 3 & $-30.1\pm 2.8$ & $2447593.33761\pm 0.00024$ & $494.85869\pm 0.00005$ & $ -30.526\pm 0.093$ &  \\
515.0 & 2 &-1 & 31 & 4 & $-22.6\pm 2.6$ & $2449065.93209\pm 0.00024$ & $515.00799\pm 0.00018$ & $ -15.661\pm 0.033$ & $6.332\pm 0.130$ \\
516.0 & 1 &+1 & 20 & 5 & $+12.8\pm 0.3$ & $2449065.93114\pm 0.00002$ & $516.06545\pm 0.00001$ & $  13.146\pm 0.003$ & $ 0.193\pm 0.008$ \\
%%516.0 & 5 & $2447593.33459\pm 0.00002$ & $516.05420\pm 0.00001$ & $  11.207\pm 0.009$ & $0.014\pm 0.008$ \\  % lower probability
517.1 & 1 & 0 & 20 & 4 & $+18.2\pm 0.8$ & $2447593.33920\pm 0.00003$ & $517.16755\pm 0.00009$ & $  15.172\pm 0.045$ & $-8.166\pm 0.273$ \\
518.2 & 1 &-1 & 20 & 3 & $ +3.2\pm 1.8$ & $2449065.93082\pm 0.00005$ & $518.29767\pm 0.00001$ & $  -0.255\pm 0.016$ &  \\
536.9 & 1 &+1 & 21 & 4 & $+28.1\pm 1.9$ & $2449065.93110\pm 0.00025$ & $536.82344\pm 0.00010$ & $  30.771\pm 0.037$ & $6.110\pm 0.071$ \\
538.1 & 1 & 0 & 21 & 3 & $ +1.0\pm 0.7$ & $2447593.33642\pm 0.00003$ & $538.15390\pm 0.00002$ & $   4.304\pm 0.010$ &  \\
539.3 & 1 &-1 & 21 & 3 & $ -1.5\pm 0.4$ & $2447593.33895\pm 0.00002$ & $539.35572\pm 0.00001$ & $  -0.339\pm 0.015$ &  \\
557.1 & 1 &+1 & 22 & 3 & $ -1.5\pm 3.3$ & $2449065.92978\pm 0.00021$ & $557.11218\pm 0.00005$ & $  -3.419\pm 0.057$ &  \\
558.4 & 1 & 0 & 22 & 3 & $ -4.3\pm 1.9$ & $2449065.92676\pm 0.00011$ & $558.44566\pm 0.00003$ & $ -10.946\pm 0.031$ &  \\
559.7 & 1 &-1 & 22 & 3 & $+12.2\pm 6.7$ & $2449065.92945\pm 0.00012$ & $559.76600\pm 0.00005$ & $  11.602\pm 0.050$ &  \\
561.9 & 2 & 0 & 41 & 3 & $-20.5\pm 3.7$ & $2449065.93013\pm 0.00034$ & $562.00642\pm 0.00004$ & $ -19.230\pm 0.044$ &  \\
731.6 & 1 &-1 & 30 & 3 & $+84.3\pm 15.$ & $2449065.92821\pm 0.00033$ & $731.61198\pm 0.00012$ & $  88.859\pm 0.142$ &  \\
819.7 & 2 &-2 & 60 & 3 & $-22.7\pm 7.8$ & $2447593.34294\pm 0.00023$ & $819.96585\pm 0.00013$ & $ -40.498\pm 0.195$ &  \\

          &        &     &     &     &                      &         &       &                                 &  \\
    \hline
  \end{tabular}
 \end{center}
}
\end{table*}
% -------------------------------------------

% ------------------------------------------------------------------------
\subsection{The 516.0 s mode}  % Sect. 3.3

%%%
The 516.0 s pulsation mode ($\ell=1$, $m=+1$, $k=20\pm2$) 
is one of the highest amplitude modes of PG~1159-035.
Using the 1983, 1985, 1989, and 1993 data sets, Costa et al. (1999) 
achieved the first direct measurement of period change in a (pre-)white dwarf, 
deriving ${\dot P}_{516} = (+ 13.0\pm 2.6) \times 10^{-11}\, {\rm ss}^{-1}$. 
They refined this value with the (second order) (O-C) method finding  
${\dot P}_{516} = (+ 13.07\pm 0.03) \times 10^{-11}\, {\rm ss}^{-1}$.

%%%
Fitting the curve $P=P_o+\dot P\, (t-T_o)$ to the data in Table~\ref{table:1},
we measure $P_o = 516.0516\pm 0.0005$ s and 
$\dot P=(+12.8\pm 0.3)\times 10^{-11}\, {\rm ss}^{-1}$
at $T_o=244\,7593.334592$ (BCT). 
Including $\ddot P$ in the fitting,
$P(T-T_o)=P_o+\dot P\cdot (T-T_o)+1/2\cdot {\ddot P}\cdot (T-T_o)^2$, we obtain
$P_o=516.0534\pm 0.0008$ s, $\dot P = (+13.2\pm 0.3)\times 10^{-11}\, {\rm ss}^{-1}$  and
$\ddot P = (-7.8\pm 0.7)\times 10^{-20} {\rm ss}^{-2}$, at the same $T_o$.
The two fitted curves are shown in Fig.~\ref{fig:1}.
Using these results as initial values in the (third order) (O-C) method, 
we find a best-fit solution with the parameters
$T_o=244\,7593.93114\pm 0.00002$ (BCT),
$P_o=516.065545\pm 0.00001$ seconds,
$\dot P= (+13.146\pm 0.003)\times 10^{-11} {\rm ss}^{-1}$, and
$\ddot P_o=(1.93\pm 0.08)\times 10^{-20} {\rm ss}^{-1}$.
With the introduction of the third order term,
the new (O-C) result for $\dot P$ differs by a margin of approximately $2.5\sigma$ 
from the previous value calculated by Costa et al. (1999).

% ----------------------------------------------------------------------------------
\subsection{The 517.1 s mode}  % Sect. 3.4

%%%
The 517.1 s period was identified as the central peak ($m=0$) of the $k=20\pm 2$ 
triplet, where the 518.2 s and 516.0 s modes are the $m=-1$ and $m=+1$ components,
respectively.
This mode appears in all the FTs, apart from the 2002 FT. 
By fitting a (linear) curve to the data in Table~\ref{table:1}, we obtain
$\dot P=(+18.2\pm 0.8) \times 10^{-11}{\rm ss}^{-1}$; 
using this value as the initial value in the third order (O-C) fitting, 
we determine as the best-fit solution
$T_o=244\,7593.33920\pm 0.00003$ (BCT), 
$P_o=517.16755\pm 0.00009$ seconds,
$\dot P_o=(+15.172\pm  0.045)\times 10^{-11}{\rm ss}^{-1}$, and 
$\ddot P_o=(-81.7\pm 2.7) \times 10^{-20}{\rm ss}^{-1}$.
The  (O-C) results differs by $\sim 3.7\sigma$ from the directly measured one.
This difference is, at least partially, due to introduction of the third order term.

% ---------------------------------------------------------------------------------------
\subsection{The 539.3 s mode}   % Sect. 3.5 
% ---------------------------------------------------------------------------------------

%%%
The 539.3 s pulsation mode, identified $\ell=1$, $m=-1$, and $k=21\pm 2$ was detected in the 1983, 1989,
and 1993 FTs only
and was the second period for which $\dot P$ was measured. 
Costa et al. (1995) used the (O-C) method applied to the same data sets to calculate its $\dot P$, 
obtaining $\dot P=(-0.82\pm 0.04)\times 10^{-11}{\rm ss}^{-1}$. 
Using the same data sets but with an improved data reduction process, and frequency determination
we obtained $\dot P=(-1.5\pm 0.4)\times 10^{-11}{\rm ss}^{-1}$
from the direct measurement and $\dot P = (-0.339\pm 0.015)\times 10^{-11}{\rm ss}^{-1}$,
$T_o=244\,7593.33895\pm 0.00002$ (BCT), and $P_o=539.35572\pm 0.00001$ seconds
from the  (O-C) fitting.
The large difference between the present (O-C) result and the (O-C) result obtained by
Costa et al. (1995) is because in this previous work 
the times of maximum and their uncertainties were calculated using a linear fitting by a single
sinusoidal curve. As demonstrated by Costa et al. (1999) and Costa \& Kepler (2000), 
the linear fitting does not take into account the interference 
of the other pulsation modes over the fitted mode and
underestimate the calculated uncertainties.

% ====================================================================================
\section{Application: calculating evolutionary timescales}\label{sect:6} % Sect. 4
% ====================================================================================

% ----------------------------------------------------------------------------------
\subsection{Variation in the stellar rotation period}\label{sect:6.1}  % Sect. 4.1
% ----------------------------------------------------------------------------------

%%%
The observed frequency spacings between the $m\ne 0$ components and 
the central  ($m=0$) peak    %%, $\nu_m - \nu_0$, 
of a multiplet are caused by a combination
of the effect of the stellar rotation and the effect of the magnetic field of the star over
the observed pulsation frequencies (see e.g. Jones et al. 1989):
\begin{equation}
 \delta \nu_m = \nu_m - \nu_0 = \delta \nu_{\rm rot,\,m} + \delta \nu_{\rm mag,\,m}\quad .
 \label{eq:8} 
\end{equation}
To a {\it first-order approximation},
the rotation splitting $\delta \nu_{\rm rot,\,m}$ is proportional to the angular
rotation frequency $\Omega_{\rm rot}$ multiplied by  $m$, while the magnetic splitting
$\delta\nu_{\rm mag}$ is proportional to the magnetic strength 
$B=|\vec{B}|$ multiplied by $m^2$ (see e.g. W91),

\begin{equation}
 \delta \nu_m \simeq m\,C\,\Omega_{\rm rot} + m^2\, D\, B^2\quad ,
  \label{eq:9}   %%%\label{eq_dnu}
\end{equation}
where $C$ and $D$ are  constants.
Costa et al. (2008) demonstrated that the PG~1159-035 magnetic strength is  weak, $B<2\,000$ G, and that the
contribution of the average magnetic splitting in the total observed splitting is lower than
1\% ($\delta\nu_{\rm mag,m}=0.007\pm 0.002\,\mu$Hz), which allows us to approximate Eq.~\ref{eq:9} with:
\begin{equation}
 \delta \nu_m \simeq m\,C\,\Omega_{\rm rot} \quad .
 \label{eq:10}   %%%\label{eq_dfo}
\end{equation}
The proportionality constant $C$ can be rewritten  (Hansen et al. 1977) as
\begin{equation}
C=1-C_0-C_1\quad ,
\label{eq:11} 
\end{equation}
where $C_0$ is the uniform rotation coefficient that depends on $k$ and $\ell$, and equals $C_0=C_0(k,\,\ell)$,
and $C_1$, related to nonuniform rotation effects, is a function of $k$, $\ell$ and $|m|$,
and equals $C_1=C_1(k,\,\ell,\,|m|)$.
The second coefficient depends on the adiabatic pulsation properties, equilibrium structure, and rotation law. 
If we assume  uniform rotation, $C_1 =0 $ and then $C\simeq 1 - C_0$.   % <--- A UNIFORM??
In the asymptotic limit with high radial overtones ($k>>1$), the uniform rotation coefficient can be 
approximated by (Brickhill 1975)  $C_0 \simeq [ \ell\, (\ell+1)  ]^{-1} $.
For $\ell=1$ modes, $C_0 \simeq 1/2$, $C \simeq 1/2$, and   Eq.~\ref{eq:10} can be rewritten as:
\begin{equation}
  \nu_m - \nu_0 \simeq m\,\frac{1}{2}\,\Omega_{\rm rot}\quad .
  \label{eq:12} 
\end{equation}
This approximation is accurate to about $5\%$ for the PG~1159-035 frequency splittings (Costa et al. 2008). 

%%%
Deriving both sides of the equation above relative to time,
\begin{equation}
  {\dot \nu}_m - {\dot \nu}_0 \simeq m\,\frac{1}{2}\,{\dot \Omega}_{\rm rot}
  \label{eq:13} 
\end{equation}
or in terms of periods,
\begin{equation}
  {\dot P}_{\rm rot} \simeq \frac{2}{m} \left( \frac{\dot{P}_m}{P_m^2} - \frac{\dot{P}_0}{P_0^2} \right)\, {P_{\rm rot}^2}\quad ,
  \label{eq:14}  %%%\label{eq_PPP}
\end{equation}
where $\dot{P}_{\rm rot}$ is the \emph{instantaneous} rotation period change rate.

%%%
Taking the time derivative of Eq.~\ref{eq:14} we obtain the second derivative $\ddot P_{\rm rot}$: 
\begin{eqnarray}
{\ddot P}_{\rm rot}  & \simeq & 
	-	\frac{4}{m}\, \left( \frac{\dot{P}_m^2}{P_m^3}  - \frac{\dot{P}_0^2}{P_0^3} \right) \,P_{\rm rot}^2
	+ 	\frac{2}{m}\, \left( \frac{\ddot{P}_m}{P_m^2} - \frac{\ddot{P}_0}{P_0^2} \right) \,P_{\rm rot}^2  + \\  \nonumber
        & + &	\frac{4}{m}\, \left( \frac{\dot{P}_m}{P_m^2} - \frac{\dot{P}_0}{P_0^2} \right) \,P_{\rm rot}\, \dot{P}_{\rm rot} \quad .\\ \nonumber
\label{eq:15}  %%%\label{eq_PPP2a}
\end{eqnarray}
In Eq.~\ref{eq:15}, the second term is the dominant term for PG~1159-035
and the other two terms can be neglected:
\begin{equation}
  {\ddot P}_{\rm rot} \simeq
	 	\frac{2}{m}\, \left( \frac{\ddot{P}_m}{P_m^2} - \frac{\ddot{P}_0}{P_0^2} \right) \,P_{\rm rot}^2 \quad .
   \label{eq:16}  %%%%\label{eq_PPP2}
\end{equation}
We note that with at least two well-determined $\dot P$ in a triplet, 
we are able to calculate the rotation period change rate,
${\dot P}_{\rm rot}$, using Eq.~\ref{eq:14}.
If $\ddot{P}_0$ and   $\ddot{P}_m$ are both known, we can also calculate $\ddot{P}_{\rm rot}$
from Eq.~\ref{eq:16}.

%%%
Using the (O-C) method results for the
($m=0$) 517.1 s and the ($m=+1$) 516.0 s modes 
and the rotation period,  $P_{\rm rot} = 1.3920\pm 0.0008$ days (Costa et al. 2008),
in  Eq.~\ref{eq:14} and  Eq.~\ref{eq:16}, we obtain:
\begin{equation}
  {\dot P}_{\rm rot} \simeq (-2.13\pm 0.05)\times 10^{-6}\,{\rm ss}^{-1} \quad  \rm{and}
  \label{eq:17}
\end{equation}
\begin{equation}
  {\ddot P}_{\rm rot} \simeq (+9.0\pm 0.3)\times 10^{-14}\,{\rm ss}^{-2} \quad  .
  \label{eq:18} 
\end{equation}    
Using the direct method results, the rates are 
${\dot P}_{\rm rot}\simeq (-2.5\pm 0.3)\times 10^{-6} \,{\rm ss}^{-1}$ 
and $ {\ddot P}_{\rm rot} \simeq (+8.0\pm 8.8)\times 10^{-14} \,{\rm ss}^{-2}$. 

%%%
The accuracy of ${\dot P}_{\rm rot}$ measured by Eq.~\ref{eq:14} 
depends strongly on the accuracies of $\dot{P}_m$ and $\dot{P}_0$, 
far more than on the  accuracies of the pulsation periods or rotation period.
Keeping unchanged $\sigma_{\dot P_0}$ and $\sigma_{\dot P_m}$ and increasing 
the uncertainties in $P_0$ and $P_m$ by a factor of 10,
the ${\dot P}_{\rm rot}$ uncertainty is almost  unaffected, 
although it changes sensitively with $\sigma_{\dot P_0}$ and $\sigma_{\dot P_m}$.
However, the uncertainties in the period determination, $\sigma_{P_m}$, 
affect indirectly the ${\dot P}_{\rm rot}$ uncertainty, 
because both $\sigma_{\dot P_0}$  and $\sigma_{\dot P_m}$ depend on  $\sigma_{P_m}$ and $\sigma_{P_0}$.
If the period uncertainty are then underestimated, 
the uncertainty in ${\dot P}_{\rm rot}$ must also be. 
The results above and the following results are discussed in the final section of this article.

% ----------------------------------------------------------
\subsection{The contraction rate}  % Sect. 4.2
% ----------------------------------------------------------

%%%
We now calculate the contraction timescale, $R/\dot R$, 
and contraction rate,
$\dot R$. For a star of uniform rotation  and negligible mass loss (Kawaler 1986; Costa et al. 1999),
\begin{equation}
  \frac{\dot R}{R}\simeq \frac{1}{2}\,\frac{{\dot P}_{\rm rot}}{P_{\rm rot}}\quad .
  \label{eq:19}  %%%\label{eq_RR} 
\end{equation}
Using the result above for $\dot P_{\rm rot}$, we obtain:
\begin{equation}
  \frac{\dot R}{R}\simeq (-8.9\pm 0.2)\times 10^{-12}\,{\rm s}^{-1}\quad ,
  \label{eq:20} 
\end{equation}
which is equivalent to 
\begin{equation}
  \frac{\dot R}{R}\simeq (-2.8\pm 0.1) \times 10^{-4}\,{\rm R}_\star/{\rm yr}\quad .
  \label{eq:21} 
\end{equation}
If we use the radius predicted by evolutionary models for PG~1159-035 by Kawaler \& Bradley (1994)
 $R_\star=(0.025\pm 0.005)\,R_\odot$, the radius change rate $\dot R$ would be:
\begin{equation}
  \dot R = (-2.2\pm 0.5)\times 10^{-13}\,R_\odot/\rm{s} 
  \label{eq:22}
\end{equation}
or
\begin{equation}
  \dot R = (-5\pm 1)\,{\rm km/yr} \quad .
  \label{eq:23} 
\end{equation}

%%%
Using an approximation of the 517.1 s mode $\dot P$ and following the
same steps, Costa et al. (1999) derived $\dot R/R \simeq (-4\pm 15)\times 10^{-11} {\rm s}^{-1}$.
With the measurement of $\dot P$ of the 517.1 s mode, the obtained value for the contraction rate is
far more constrained, but, 
as pointed out by the authors, more realistic estimations of $\dot R/R$ must use {\it differential
rotation} at least for the outer layer of the star. Models for differentially rotating white dwarf
stars calculated by Ostriker \& Bondenheimer (1968) suggest that the center rotates more rapidly
than the outer layers, but not significantly so i.e. $\Omega_{\rm surface} /\Omega_{\rm center} > 0.2$. 
In this case, our $\dot R/R$ calculated from Eq.\ref{eq:19} must be seen as an upper limit for the actual value:
\begin{equation}
  \left| \frac{\dot R}{R}  \right| < -8.9\times 10^{-12}\,{\rm s}^{-1}\quad .
  \label{eq:24} 
\end{equation}

% -------------------------------------------------------------------------------------
\subsection{The cooling rate}  % Sect. 4.3
% -------------------------------------------------------------------------------------

%%%
We can now obtain a first estimate for the PG~1159-035 cooling rate, $\dot T$. 
The changes in period are related to two physical processes in the star:  
the cooling of the star and the envelope contraction
(see e.g. Winget et al. 1983; Kawaler et al. 1985a):
\begin{equation}
  \frac{\dot P}{P} \simeq -a\, \frac{\dot{T}_m}{T_m} + b\, \frac{\dot R}{R}
  \label{eq:25}  %%%\label{eq_PTR}
\end{equation}
where $P$ is the pulsation period (for the $m=0$ multiplet component),
$T_m$ is the temperature at the region where the model's weight function
has maximum weight (where we emphasize that $T_m$ is the temperature, not the time of maximum, 
$T_{\rm max}$),  $R$ is the stellar radius, and $\dot P$, $\dot{T}_m$, and $\dot R$, are
the respective temporal variation rates. 
The constants $a$ and $b$ are positive numbers and, roughly, $a\simeq 1/2$ and $b\simeq 1$
(Kawaler et al. 1985b). Then,
\begin{equation}
  \frac{\dot{T}_m}{T_m} = 2\,\left( - \frac{\dot P}{P} + \frac{\dot R}{R} \right)\quad .
  \label{eq:26} 
\end{equation} 
Using $\dot P/P=+2.94\times 10^{-13}\, {\rm s}^{-1}$ for the ($m=0$) 517.0 s pulsation mode, we obtain
\begin{equation}
  \frac{\dot{T}_m}{T_m} =( -1.84\pm 0.04)\times 10^{-11}\, {\rm s}^{-1} \quad .   % << RECALCULAR SIGMA....
  \label{eq:27} 
\end{equation}
Since $\dot R/R=0.89\times 10^{-11}\,{\rm s}{-1}$, Eq.~\ref{eq:25} implies that the
PG~1159-035 temporal change in period is controlled far more by the stellar cooling than by the contraction of its
envelope, as  expected. The evolution of  white dwarf and pre-white dwarf stars is dominated by cooling,
but in hot pre-white dwarfs, the envelope contraction is still significant.

%%%
In bright pulsators such as PG~1159-035  ($\,\rm{Log}(L/L_{\odot}) \simeq 2.6$), 
the weight function reaches a maximum closer to the core of the star  
(Kawaler et al. 1985b). 
If we assume $T_m \simeq T_{\rm core} \simeq 7.72\times 10^7$ K (Alejandro C\'orsico, personal communication),
\begin{equation}
  \dot{T}_m \simeq -1.42\times 10^{-3}\,{\rm K/s} \simeq -45\,000\,{\rm K/yr} \quad .
  \label{eq:28} 
\end{equation}

% ===================================================

% ===================================================
\section{Summary and discussion} % Sect. 5
% ===================================================

%%%
The main results of our present work are:

%%%
\begin{enumerate}  

\item   The PG~1159-035 pulsation periods vary at rates of between 0.5 and 1.0 ms/years.
        For the WET data sets, the accuracy in the period determination is 10 ms for the least accurate
        cases (low amplitudes) and 1 ms for the most accurate ones, which enables the direct measurement of the pulsation
        periods after a few years of observation. After 19 years of observations, the $\dot P$
        of the 516.0 s mode (one of the modes with higher amplitude) was measured with
        a relative standard uncertainty ($\sigma_{\dot P}/\,|{\dot P}|$ of 2\% and its $\ddot P$ 
        could be estimated with  relative standard uncertainty of 10\%.

\item   The $\dot P$ of the 27 periods present in three or more yearly Fourier transforms were 
        measured directly and the  values were 
        refined using  (O-C) fitting. For seven of the pulsation modes, we used  third-order
	(O-C) fitting to calculate  $\ddot P$. The measured $|\ddot P|$ are between 
        $1.4\times 10^{-21}\,{\rm ss}^{-2}$ and
	$8.7\times 10^{-19}\,{\rm ss}^{-2}$.

\item   Using the $\dot P$ of the $m=0$ and $m=+1$ components of the
	517.1 s multiplet, we estimated the rotation period change to be
	$\dot P_{\rm rot}=(-2.13\pm 0.05)\times 10^{-6}\,{\rm ss}^{-1}$ or
	67.2 s/yr.

\item   From the $\ddot P$ of the same multiplet, we estimated that the second order
        variation in the rotational period was $\ddot P = (+9.0\pm 0.3)\times 10^{-14}\,{\rm ss}^{-2}$. 

\item   If we used the calculated values for $\dot P_{\rm rot}$, the PG~1159-035
        contraction rate was $\dot R \simeq (-5\pm 1)$ km/s for $R_\star \simeq (0.025\pm 0.005)\,R_\odot$,
        assuming uniform rotation, and the cooling rate is $\dot{T}_m \simeq -45\,000$ K/yr,
        assuming $T_m\simeq T_{\rm core} \simeq 7.72\times 10^7$ K. 

\end{enumerate}

%%%
Our results demonstrate that some periods decrease, while others increase, at different rates,
even for the same multiplet components.
Evolutionary models calculated by Kawaler \& Bradley (1994) and  the La Plata group
(C\'orsico et al. 2008; Althaus et al. 2008) 
predict that the $\dot P$ of ($\ell=1$ and $m=0$) pulsation modes  must have different values 
and signs (some increase, while other decrease) due to the trapping of pulsation
modes in different  layers of the star, but the calculated values do not fit
the observed $\dot P$s. 
The interior, evolution, and pulsation mechanisms in PG~1159 stars are not well known. 
New models have been developed by the La Plata group to fit the observed results
and will be published in a future article.

%%%
A negative value for ${\dot P}_{\rm rot}$ is expected, because 
pre-white dwarf stars such as PG~1159-035 undergo rapid envelope contraction processes. 
With contraction, the stellar radius decreases. 
To conserve angular momentum, the angular rotation speed must increase. 
In other words, its rotation period decreases, and therefore ${\dot P}_{\rm rot}<0$.
The shorter rotation periods observed in white dwarfs are of the order of few hours.
This is the case, for example, for the DBV \object{EC20058-5234} ($P_{\rm rot} \simeq 2$ h, 
Sullivan et al. 2008) and of the magnetic \object{GD~356}  ($P_{\rm rot} = 2.6$ h,
Brinkworth et al. 2004). If the PG~1159-035 rotation period changed from its current
value to $P_{\rm rot} = 2$ h during the DOV phase ($\sim 10^6$ yr), the variation rate would be 
${\dot P}_{\rm rot} \simeq -0.1$ s/yr.
However, the calculations in Sect.~\ref{sect:6.1} estimate that the rotation period is changing 
at a ${\dot P}_{\rm rot}\simeq (-67\pm 2)\, \rm{s/year}$ rate, far higher than expected
and ${\ddot P}_{\rm rot}$ is too excessively high to be physically acceptable. 
A possible explanation is that the second order effects of the stellar rotation over
the rotational splitting are non-negligible as assumed in Eq.~\ref{eq:9}. 
The simplification of assuming a uniform rotation is also an error source in the
${\dot P}_{\rm rot}$ calculation, but the difference in the final result  must not be larger than
a factor of 10 (Ostriker \& Bodenheimer 1968).
On the other hand, a high de/acceleration of the stellar rotation is
expected if the star recently experienced a late thermal pulse (due to ejection of  matter
from the star) and this may be the case for PG~1159-035. 
We note that if $|{\dot P}_{\rm rot}|$ is overestimated, 
both the contraction rate $|\dot R|$ and the cooling rate $|\dot T|$ must 
also be overestimated.

%%% 
The accuracy in the determination of ${\dot P}_{\rm rot}$ by Eq.~\ref{eq:14} 
depends strongly on the accuracies of $\dot{P}_m$ and $\dot{P}_0$, 
far more than on the period accuracies.

%%%
Although these first results are not fully understood, 
we have shown that several long-term campaigns appear to be reaching sufficient 
accuracy to interpret evolutionary changes. 
The last PG~1159-035 observational campaigns were carried out in 2002 by WET. 
New observations of PG~1159-035 in future years should allow
the direct measurement of additional pulsation modes $\dot P$ and $\ddot P$ 
and the refining of derived values to date, 
improving the calculation of the PG~1159-035 evolutionary rates.

% --------------------------------------------------------------------------------------
% ---------------------------------------------------------------------------------------
\begin{acknowledgements}
      This work was partially supported by CNPq - Brazil.
      We thanks Alejandro C\'orsico for the calculation of the PG~1159-035
      $T_{\rm core}$. 
\end{acknowledgements}

\end{document}